\documentclass[12pt,prd,showpacs,tightenlines,nofootinbib]{revtex4}
\usepackage{bm}
\usepackage{graphics}
\usepackage{rotating}
\usepackage{epsfig}

\begin{document}

\begin{flushright}
Version 2
\end{flushright}

\title{MUONIC HYDROGEN GROUND STATE \\HYPERFINE SPLITTING
\footnote{Talk presented at the 11 th Lomonosov Conference on
Elementary Particle Physics, Moscow State University, August 2003}}

\author{R.N.Faustov\footnote{faustov@theory.sinp.msu.ru}}

\affiliation{Russian Academy of Sciences, Scientific Council
for Cybernetics, Vavilov Street 40, 117333 Moscow, Russia}

\author{A.P.Martynenko\footnote{apm@physik.hu-berlin.de}}

\address{Samara State University, Theoretical Physics
Department, Pavlov Street 1, 443011 Samara, Russia}

\begin{abstract}
Corrections of orders $\alpha^5$, $\alpha^6$ are calculated in the hyperfine
splitting of the muonic hydrogen ground state. The nuclear structure effects
are taken into account in the one- and two-loop
Feynman amplitudes by means of the proton electromagnetic form factors.
The modification of the
hyperfine splitting part of the Breit potential due to the electron vacuum
polarization is considered.
Total numerical value of the $1S$-state hyperfine splitting 182.638 meV in the
$\mu p$ can play the role of
proper estimation for the corresponding experiment with the accuracy $30$ ppm.
\end{abstract}


\maketitle

\immediate\write16{<<WARNING: LINEDRAW macros work with emTeX-dvivers
                    and other drivers supporting emTeX \special's
                    (dviscr, dvihplj, dvidot, dvips, dviwin, etc.) >>}

\newdimen\Lengthunit       \Lengthunit  = 1.5cm
\newcount\Nhalfperiods     \Nhalfperiods= 9
\newcount\magnitude        \magnitude = 1000

\catcode`\*=11
\newdimen\L*   \newdimen\d*   \newdimen\d**
\newdimen\dm*  \newdimen\dd*  \newdimen\dt*
\newdimen\a*   \newdimen\b*   \newdimen\c*
\newdimen\a**  \newdimen\b**
\newdimen\xL*  \newdimen\yL*
\newdimen\rx*  \newdimen\ry*
\newdimen\tmp* \newdimen\linwid*

\newcount\k*   \newcount\l*   \newcount\m*
\newcount\k**  \newcount\l**  \newcount\m**
\newcount\n*   \newcount\dn*  \newcount\r*
\newcount\N*   \newcount\*one \newcount\*two  \*one=1 \*two=2
\newcount\*ths \*ths=1000
\newcount\angle*  \newcount\q*  \newcount\q**
\newcount\angle** \angle**=0
\newcount\sc*     \sc*=0

\newtoks\cos*  \cos*={1}
\newtoks\sin*  \sin*={0}

\catcode`\[=13

\def\rotate(#1){\advance\angle**#1\angle*=\angle**
\q**=\angle*\ifnum\q**<0\q**=-\q**\fi
\ifnum\q**>360\q*=\angle*\divide\q*360\multiply\q*360\advance\angle*-\q*\fi
\ifnum\angle*<0\advance\angle*360\fi\q**=\angle*\divide\q**90\q**=\q**
\def\sgcos*{+}\def\sgsin*{+}\relax
\ifcase\q**\or
 \def\sgcos*{-}\def\sgsin*{+}\or
 \def\sgcos*{-}\def\sgsin*{-}\or
 \def\sgcos*{+}\def\sgsin*{-}\else\fi
\q*=\q**
\multiply\q*90\advance\angle*-\q*
\ifnum\angle*>45\sc*=1\angle*=-\angle*\advance\angle*90\else\sc*=0\fi
\def[##1,##2]{\ifnum\sc*=0\relax
\edef\cs*{\sgcos*.##1}\edef\sn*{\sgsin*.##2}\ifcase\q**\or
 \edef\cs*{\sgcos*.##2}\edef\sn*{\sgsin*.##1}\or
 \edef\cs*{\sgcos*.##1}\edef\sn*{\sgsin*.##2}\or
 \edef\cs*{\sgcos*.##2}\edef\sn*{\sgsin*.##1}\else\fi\else
\edef\cs*{\sgcos*.##2}\edef\sn*{\sgsin*.##1}\ifcase\q**\or
 \edef\cs*{\sgcos*.##1}\edef\sn*{\sgsin*.##2}\or
 \edef\cs*{\sgcos*.##2}\edef\sn*{\sgsin*.##1}\or
 \edef\cs*{\sgcos*.##1}\edef\sn*{\sgsin*.##2}\else\fi\fi
\cos*={\cs*}\sin*={\sn*}\global\edef\gcos*{\cs*}\global\edef\gsin*{\sn*}}\relax
\ifcase\angle*[9999,0]\or
[999,017]\or[999,034]\or[998,052]\or[997,069]\or[996,087]\or
[994,104]\or[992,121]\or[990,139]\or[987,156]\or[984,173]\or
[981,190]\or[978,207]\or[974,224]\or[970,241]\or[965,258]\or
[961,275]\or[956,292]\or[951,309]\or[945,325]\or[939,342]\or
[933,358]\or[927,374]\or[920,390]\or[913,406]\or[906,422]\or
[898,438]\or[891,453]\or[882,469]\or[874,484]\or[866,499]\or
[857,515]\or[848,529]\or[838,544]\or[829,559]\or[819,573]\or
[809,587]\or[798,601]\or[788,615]\or[777,629]\or[766,642]\or
[754,656]\or[743,669]\or[731,681]\or[719,694]\or[707,707]\or
\else[9999,0]\fi}

\catcode`\[=12

\def\GRAPH(hsize=#1)#2{\hbox to #1\Lengthunit{#2\hss}}

\def\Linewidth#1{\global\linwid*=#1\relax
\global\divide\linwid*10\global\multiply\linwid*\mag
\global\divide\linwid*100\special{em:linewidth \the\linwid*}}

\Linewidth{.4pt}
\def\sm*{\special{em:moveto}}
\def\sl*{\special{em:lineto}}
\let\moveto=\sm*
\let\lineto=\sl*
\newbox\spm*   \newbox\spl*
\setbox\spm*\hbox{\sm*}
\setbox\spl*\hbox{\sl*}

\def\mov#1(#2,#3)#4{\rlap{\L*=#1\Lengthunit
\xL*=#2\L* \yL*=#3\L*
\xL*=\xscale\xL* \yL*=\yscale\yL*
\rx* \the\cos*\xL* \tmp* \the\sin*\yL* \advance\rx*-\tmp*
\ry* \the\cos*\yL* \tmp* \the\sin*\xL* \advance\ry*\tmp*
\kern\rx*\raise\ry*\hbox{#4}}}

\def\rmov*(#1,#2)#3{\rlap{\xL*=#1\yL*=#2\relax
\rx* \the\cos*\xL* \tmp* \the\sin*\yL* \advance\rx*-\tmp*
\ry* \the\cos*\yL* \tmp* \the\sin*\xL* \advance\ry*\tmp*
\kern\rx*\raise\ry*\hbox{#3}}}

\def\lin#1(#2,#3){\rlap{\sm*\mov#1(#2,#3){\sl*}}}

\def\arr*(#1,#2,#3){\rmov*(#1\dd*,#1\dt*){\sm*
\rmov*(#2\dd*,#2\dt*){\rmov*(#3\dt*,-#3\dd*){\sl*}}\sm*
\rmov*(#2\dd*,#2\dt*){\rmov*(-#3\dt*,#3\dd*){\sl*}}}}

\def\arrow#1(#2,#3){\rlap{\lin#1(#2,#3)\mov#1(#2,#3){\relax
\d**=-.012\Lengthunit\dd*=#2\d**\dt*=#3\d**
\arr*(1,10,4)\arr*(3,8,4)\arr*(4.8,4.2,3)}}}

\def\arrlin#1(#2,#3){\rlap{\L*=#1\Lengthunit\L*=.5\L*
\lin#1(#2,#3)\rmov*(#2\L*,#3\L*){\arrow.1(#2,#3)}}}

\def\dasharrow#1(#2,#3){\rlap{{\Lengthunit=0.9\Lengthunit
\dashlin#1(#2,#3)\mov#1(#2,#3){\sm*}}\mov#1(#2,#3){\sl*
\d**=-.012\Lengthunit\dd*=#2\d**\dt*=#3\d**
\arr*(1,10,4)\arr*(3,8,4)\arr*(4.8,4.2,3)}}}

\def\clap#1{\hbox to 0pt{\hss #1\hss}}

\def\ind(#1,#2)#3{\rlap{\L*=.1\Lengthunit
\xL*=#1\L* \yL*=#2\L*
\rx* \the\cos*\xL* \tmp* \the\sin*\yL* \advance\rx*-\tmp*
\ry* \the\cos*\yL* \tmp* \the\sin*\xL* \advance\ry*\tmp*
\kern\rx*\raise\ry*\hbox{\lower2pt\clap{$#3$}}}}

\def\sh*(#1,#2)#3{\rlap{\dm*=\the\n*\d**
\xL*=\xscale\dm* \yL*=\yscale\dm* \xL*=#1\xL* \yL*=#2\yL*
\rx* \the\cos*\xL* \tmp* \the\sin*\yL* \advance\rx*-\tmp*
\ry* \the\cos*\yL* \tmp* \the\sin*\xL* \advance\ry*\tmp*
\kern\rx*\raise\ry*\hbox{#3}}}

\def\calcnum*#1(#2,#3){\a*=1000sp\b*=1000sp\a*=#2\a*\b*=#3\b*
\ifdim\a*<0pt\a*-\a*\fi\ifdim\b*<0pt\b*-\b*\fi
\ifdim\a*>\b*\c*=.96\a*\advance\c*.4\b*
\else\c*=.96\b*\advance\c*.4\a*\fi
\k*\a*\multiply\k*\k*\l*\b*\multiply\l*\l*
\m*\k*\advance\m*\l*\n*\c*\r*\n*\multiply\n*\n*
\dn*\m*\advance\dn*-\n*\divide\dn*2\divide\dn*\r*
\advance\r*\dn*
\c*=\the\Nhalfperiods5sp\c*=#1\c*\ifdim\c*<0pt\c*-\c*\fi
\multiply\c*\r*\N*\c*\divide\N*10000}

\def\dashlin#1(#2,#3){\rlap{\calcnum*#1(#2,#3)\relax
\d**=#1\Lengthunit\ifdim\d**<0pt\d**-\d**\fi
\divide\N*2\multiply\N*2\advance\N*\*one
\divide\d**\N*\sm*\n*\*one\sh*(#2,#3){\sl*}\loop
\advance\n*\*one\sh*(#2,#3){\sm*}\advance\n*\*one
\sh*(#2,#3){\sl*}\ifnum\n*<\N*\repeat}}

\def\dashdotlin#1(#2,#3){\rlap{\calcnum*#1(#2,#3)\relax
\d**=#1\Lengthunit\ifdim\d**<0pt\d**-\d**\fi
\divide\N*2\multiply\N*2\advance\N*1\multiply\N*2\relax
\divide\d**\N*\sm*\n*\*two\sh*(#2,#3){\sl*}\loop
\advance\n*\*one\sh*(#2,#3){\kern-1.48pt\lower.5pt\hbox{\rm.}}\relax
\advance\n*\*one\sh*(#2,#3){\sm*}\advance\n*\*two
\sh*(#2,#3){\sl*}\ifnum\n*<\N*\repeat}}

\def\shl*(#1,#2)#3{\kern#1#3\lower#2#3\hbox{\unhcopy\spl*}}

\def\trianglin#1(#2,#3){\rlap{\toks0={#2}\toks1={#3}\calcnum*#1(#2,#3)\relax
\dd*=.57\Lengthunit\dd*=#1\dd*\divide\dd*\N*
\divide\dd*\*ths \multiply\dd*\magnitude
\d**=#1\Lengthunit\ifdim\d**<0pt\d**-\d**\fi
\multiply\N*2\divide\d**\N*\sm*\n*\*one\loop
\shl**{\dd*}\dd*-\dd*\advance\n*2\relax
\ifnum\n*<\N*\repeat\n*\N*\shl**{0pt}}}

\def\wavelin#1(#2,#3){\rlap{\toks0={#2}\toks1={#3}\calcnum*#1(#2,#3)\relax
\dd*=.23\Lengthunit\dd*=#1\dd*\divide\dd*\N*
\divide\dd*\*ths \multiply\dd*\magnitude
\d**=#1\Lengthunit\ifdim\d**<0pt\d**-\d**\fi
\multiply\N*4\divide\d**\N*\sm*\n*\*one\loop
\shl**{\dd*}\dt*=1.3\dd*\advance\n*\*one
\shl**{\dt*}\advance\n*\*one
\shl**{\dd*}\advance\n*\*two
\dd*-\dd*\ifnum\n*<\N*\repeat\n*\N*\shl**{0pt}}}

\def\w*lin(#1,#2){\rlap{\toks0={#1}\toks1={#2}\d**=\Lengthunit\dd*=-.12\d**
\divide\dd*\*ths \multiply\dd*\magnitude
\N*8\divide\d**\N*\sm*\n*\*one\loop
\shl**{\dd*}\dt*=1.3\dd*\advance\n*\*one
\shl**{\dt*}\advance\n*\*one
\shl**{\dd*}\advance\n*\*one
\shl**{0pt}\dd*-\dd*\advance\n*1\ifnum\n*<\N*\repeat}}

\def\l*arc(#1,#2)[#3][#4]{\rlap{\toks0={#1}\toks1={#2}\d**=\Lengthunit
\dd*=#3.037\d**\dd*=#4\dd*\dt*=#3.049\d**\dt*=#4\dt*\ifdim\d**>10mm\relax
\d**=.25\d**\n*\*one\shl**{-\dd*}\n*\*two\shl**{-\dt*}\n*3\relax
\shl**{-\dd*}\n*4\relax\shl**{0pt}\else
\ifdim\d**>5mm\d**=.5\d**\n*\*one\shl**{-\dt*}\n*\*two
\shl**{0pt}\else\n*\*one\shl**{0pt}\fi\fi}}

\def\d*arc(#1,#2)[#3][#4]{\rlap{\toks0={#1}\toks1={#2}\d**=\Lengthunit
\dd*=#3.037\d**\dd*=#4\dd*\d**=.25\d**\sm*\n*\*one\shl**{-\dd*}\relax
\n*3\relax\sh*(#1,#2){\xL*=\xscale\dd*\yL*=\yscale\dd*
\kern#2\xL*\lower#1\yL*\hbox{\sm*}}\n*4\relax\shl**{0pt}}}

\def\shl**#1{\c*=\the\n*\d**\d*=#1\relax
\a*=\the\toks0\c*\b*=\the\toks1\d*\advance\a*-\b*
\b*=\the\toks1\c*\d*=\the\toks0\d*\advance\b*\d*
\a*=\xscale\a*\b*=\yscale\b*
\rx* \the\cos*\a* \tmp* \the\sin*\b* \advance\rx*-\tmp*
\ry* \the\cos*\b* \tmp* \the\sin*\a* \advance\ry*\tmp*
\raise\ry*\rlap{\kern\rx*\unhcopy\spl*}}

\def\wlin*#1(#2,#3)[#4]{\rlap{\toks0={#2}\toks1={#3}\relax
\c*=#1\l*\c*\c*=.01\Lengthunit\m*\c*\divide\l*\m*
\c*=\the\Nhalfperiods5sp\multiply\c*\l*\N*\c*\divide\N*\*ths
\divide\N*2\multiply\N*2\advance\N*\*one
\dd*=.002\Lengthunit\dd*=#4\dd*\multiply\dd*\l*\divide\dd*\N*
\divide\dd*\*ths \multiply\dd*\magnitude
\d**=#1\multiply\N*4\divide\d**\N*\sm*\n*\*one\loop
\shl**{\dd*}\dt*=1.3\dd*\advance\n*\*one
\shl**{\dt*}\advance\n*\*one
\shl**{\dd*}\advance\n*\*two
\dd*-\dd*\ifnum\n*<\N*\repeat\n*\N*\shl**{0pt}}}

\def\wavebox#1{\setbox0\hbox{#1}\relax
\a*=\wd0\advance\a*14pt\b*=\ht0\advance\b*\dp0\advance\b*14pt\relax
\hbox{\kern9pt\relax
\rmov*(0pt,\ht0){\rmov*(-7pt,7pt){\wlin*\a*(1,0)[+]\wlin*\b*(0,-1)[-]}}\relax
\rmov*(\wd0,-\dp0){\rmov*(7pt,-7pt){\wlin*\a*(-1,0)[+]\wlin*\b*(0,1)[-]}}\relax
\box0\kern9pt}}

\def\rectangle#1(#2,#3){\relax
\lin#1(#2,0)\lin#1(0,#3)\mov#1(0,#3){\lin#1(#2,0)}\mov#1(#2,0){\lin#1(0,#3)}}

\def\dashrectangle#1(#2,#3){\dashlin#1(#2,0)\dashlin#1(0,#3)\relax
\mov#1(0,#3){\dashlin#1(#2,0)}\mov#1(#2,0){\dashlin#1(0,#3)}}

\def\waverectangle#1(#2,#3){\L*=#1\Lengthunit\a*=#2\L*\b*=#3\L*
\ifdim\a*<0pt\a*-\a*\def\x*{-1}\else\def\x*{1}\fi
\ifdim\b*<0pt\b*-\b*\def\y*{-1}\else\def\y*{1}\fi
\wlin*\a*(\x*,0)[-]\wlin*\b*(0,\y*)[+]\relax
\mov#1(0,#3){\wlin*\a*(\x*,0)[+]}\mov#1(#2,0){\wlin*\b*(0,\y*)[-]}}

\def\calcparab*{\ifnum\n*>\m*\k*\N*\advance\k*-\n*\else\k*\n*\fi
\a*=\the\k* sp\a*=10\a*\b*\dm*\advance\b*-\a*\k*\b*
\a*=\the\*ths\b*\divide\a*\l*\multiply\a*\k*
\divide\a*\l*\k*\*ths\r*\a*\advance\k*-\r*\dt*=\the\k*\L*}

\def\arcto#1(#2,#3)[#4]{\rlap{\toks0={#2}\toks1={#3}\calcnum*#1(#2,#3)\relax
\dm*=135sp\dm*=#1\dm*\d**=#1\Lengthunit\ifdim\dm*<0pt\dm*-\dm*\fi
\multiply\dm*\r*\a*=.3\dm*\a*=#4\a*\ifdim\a*<0pt\a*-\a*\fi
\advance\dm*\a*\N*\dm*\divide\N*10000\relax
\divide\N*2\multiply\N*2\advance\N*\*one
\L*=-.25\d**\L*=#4\L*\divide\d**\N*\divide\L*\*ths
\m*\N*\divide\m*2\dm*=\the\m*5sp\l*\dm*\sm*\n*\*one\loop
\calcparab*\shl**{-\dt*}\advance\n*1\ifnum\n*<\N*\repeat}}

\def\arrarcto#1(#2,#3)[#4]{\L*=#1\Lengthunit\L*=.54\L*
\arcto#1(#2,#3)[#4]\rmov*(#2\L*,#3\L*){\d*=.457\L*\d*=#4\d*\d**-\d*
\rmov*(#3\d**,#2\d*){\arrow.02(#2,#3)}}}

\def\dasharcto#1(#2,#3)[#4]{\rlap{\toks0={#2}\toks1={#3}\relax
\calcnum*#1(#2,#3)\dm*=\the\N*5sp\a*=.3\dm*\a*=#4\a*\ifdim\a*<0pt\a*-\a*\fi
\advance\dm*\a*\N*\dm*
\divide\N*20\multiply\N*2\advance\N*1\d**=#1\Lengthunit
\L*=-.25\d**\L*=#4\L*\divide\d**\N*\divide\L*\*ths
\m*\N*\divide\m*2\dm*=\the\m*5sp\l*\dm*
\sm*\n*\*one\loop\calcparab*
\shl**{-\dt*}\advance\n*1\ifnum\n*>\N*\else\calcparab*
\sh*(#2,#3){\xL*=#3\dt* \yL*=#2\dt*
\rx* \the\cos*\xL* \tmp* \the\sin*\yL* \advance\rx*\tmp*
\ry* \the\cos*\yL* \tmp* \the\sin*\xL* \advance\ry*-\tmp*
\kern\rx*\lower\ry*\hbox{\sm*}}\fi
\advance\n*1\ifnum\n*<\N*\repeat}}

\def\*shl*#1{\c*=\the\n*\d**\advance\c*#1\a**\d*\dt*\advance\d*#1\b**
\a*=\the\toks0\c*\b*=\the\toks1\d*\advance\a*-\b*
\b*=\the\toks1\c*\d*=\the\toks0\d*\advance\b*\d*
\rx* \the\cos*\a* \tmp* \the\sin*\b* \advance\rx*-\tmp*
\ry* \the\cos*\b* \tmp* \the\sin*\a* \advance\ry*\tmp*
\raise\ry*\rlap{\kern\rx*\unhcopy\spl*}}

\def\calcnormal*#1{\b**=10000sp\a**\b**\k*\n*\advance\k*-\m*
\multiply\a**\k*\divide\a**\m*\a**=#1\a**\ifdim\a**<0pt\a**-\a**\fi
\ifdim\a**>\b**\d*=.96\a**\advance\d*.4\b**
\else\d*=.96\b**\advance\d*.4\a**\fi
\d*=.01\d*\r*\d*\divide\a**\r*\divide\b**\r*
\ifnum\k*<0\a**-\a**\fi\d*=#1\d*\ifdim\d*<0pt\b**-\b**\fi
\k*\a**\a**=\the\k*\dd*\k*\b**\b**=\the\k*\dd*}

\def\wavearcto#1(#2,#3)[#4]{\rlap{\toks0={#2}\toks1={#3}\relax
\calcnum*#1(#2,#3)\c*=\the\N*5sp\a*=.4\c*\a*=#4\a*\ifdim\a*<0pt\a*-\a*\fi
\advance\c*\a*\N*\c*\divide\N*20\multiply\N*2\advance\N*-1\multiply\N*4\relax
\d**=#1\Lengthunit\dd*=.012\d**
\divide\dd*\*ths \multiply\dd*\magnitude
\ifdim\d**<0pt\d**-\d**\fi\L*=.25\d**
\divide\d**\N*\divide\dd*\N*\L*=#4\L*\divide\L*\*ths
\m*\N*\divide\m*2\dm*=\the\m*0sp\l*\dm*
\sm*\n*\*one\loop\calcnormal*{#4}\calcparab*
\*shl*{1}\advance\n*\*one\calcparab*
\*shl*{1.3}\advance\n*\*one\calcparab*
\*shl*{1}\advance\n*2\dd*-\dd*\ifnum\n*<\N*\repeat\n*\N*\shl**{0pt}}}

\def\triangarcto#1(#2,#3)[#4]{\rlap{\toks0={#2}\toks1={#3}\relax
\calcnum*#1(#2,#3)\c*=\the\N*5sp\a*=.4\c*\a*=#4\a*\ifdim\a*<0pt\a*-\a*\fi
\advance\c*\a*\N*\c*\divide\N*20\multiply\N*2\advance\N*-1\multiply\N*2\relax
\d**=#1\Lengthunit\dd*=.012\d**
\divide\dd*\*ths \multiply\dd*\magnitude
\ifdim\d**<0pt\d**-\d**\fi\L*=.25\d**
\divide\d**\N*\divide\dd*\N*\L*=#4\L*\divide\L*\*ths
\m*\N*\divide\m*2\dm*=\the\m*0sp\l*\dm*
\sm*\n*\*one\loop\calcnormal*{#4}\calcparab*
\*shl*{1}\advance\n*2\dd*-\dd*\ifnum\n*<\N*\repeat\n*\N*\shl**{0pt}}}

\def\hr*#1{\L*=\xscale\Lengthunit\ifnum
\angle**=0\clap{\vrule width#1\L* height.1pt}\else
\L*=#1\L*\L*=.5\L*\rmov*(-\L*,0pt){\sm*}\rmov*(\L*,0pt){\sl*}\fi}

\def\shade#1[#2]{\rlap{\Lengthunit=#1\Lengthunit
\special{em:linewidth .001pt}\relax
\mov(0,#2.05){\hr*{.994}}\mov(0,#2.1){\hr*{.980}}\relax
\mov(0,#2.15){\hr*{.953}}\mov(0,#2.2){\hr*{.916}}\relax
\mov(0,#2.25){\hr*{.867}}\mov(0,#2.3){\hr*{.798}}\relax
\mov(0,#2.35){\hr*{.715}}\mov(0,#2.4){\hr*{.603}}\relax
\mov(0,#2.45){\hr*{.435}}\special{em:linewidth \the\linwid*}}}

\def\dshade#1[#2]{\rlap{\special{em:linewidth .001pt}\relax
\Lengthunit=#1\Lengthunit\if#2-\def\t*{+}\else\def\t*{-}\fi
\mov(0,\t*.025){\relax
\mov(0,#2.05){\hr*{.995}}\mov(0,#2.1){\hr*{.988}}\relax
\mov(0,#2.15){\hr*{.969}}\mov(0,#2.2){\hr*{.937}}\relax
\mov(0,#2.25){\hr*{.893}}\mov(0,#2.3){\hr*{.836}}\relax
\mov(0,#2.35){\hr*{.760}}\mov(0,#2.4){\hr*{.662}}\relax
\mov(0,#2.45){\hr*{.531}}\mov(0,#2.5){\hr*{.320}}\relax
\special{em:linewidth \the\linwid*}}}}

\def\vdot{\rlap{\kern-1.9pt\lower1.8pt\hbox{$\scriptstyle\bullet$}}}
\def\vtimes{\rlap{\kern-3pt\lower1.8pt\hbox{$\scriptstyle\times$}}}
\def\vDot{\rlap{\kern-2.3pt\lower2.7pt\hbox{$\bullet$}}}
\def\vTimes{\rlap{\kern-3.6pt\lower2.4pt\hbox{$\times$}}}

\def\arc(#1)[#2,#3]{{\k*=#2\l*=#3\m*=\l*
\advance\m*-6\ifnum\k*>\l*\relax\else
{\rotate(#2)\mov(#1,0){\sm*}}\loop
\ifnum\k*<\m*\advance\k*5{\rotate(\k*)\mov(#1,0){\sl*}}\repeat
{\rotate(#3)\mov(#1,0){\sl*}}\fi}}

\def\dasharc(#1)[#2,#3]{{\k**=#2\n*=#3\advance\n*-1\advance\n*-\k**
\L*=1000sp\L*#1\L* \multiply\L*\n* \multiply\L*\Nhalfperiods
\divide\L*57\N*\L* \divide\N*2000\ifnum\N*=0\N*1\fi
\r*\n*  \divide\r*\N* \ifnum\r*<2\r*2\fi
\m**\r* \divide\m**2 \l**\r* \advance\l**-\m** \N*\n* \divide\N*\r*
\k**\r* \multiply\k**\N* \dn*\n* \advance\dn*-\k** \divide\dn*2\advance\dn*\*one
\r*\l** \divide\r*2\advance\dn*\r* \advance\N*-2\k**#2\relax
\ifnum\l**<6{\rotate(#2)\mov(#1,0){\sm*}}\advance\k**\dn*
{\rotate(\k**)\mov(#1,0){\sl*}}\advance\k**\m**
{\rotate(\k**)\mov(#1,0){\sm*}}\loop
\advance\k**\l**{\rotate(\k**)\mov(#1,0){\sl*}}\advance\k**\m**
{\rotate(\k**)\mov(#1,0){\sm*}}\advance\N*-1\ifnum\N*>0\repeat
{\rotate(#3)\mov(#1,0){\sl*}}\else\advance\k**\dn*
\arc(#1)[#2,\k**]\loop\advance\k**\m** \r*\k**
\advance\k**\l** {\arc(#1)[\r*,\k**]}\relax
\advance\N*-1\ifnum\N*>0\repeat
\advance\k**\m**\arc(#1)[\k**,#3]\fi}}

\def\triangarc#1(#2)[#3,#4]{{\k**=#3\n*=#4\advance\n*-\k**
\L*=1000sp\L*#2\L* \multiply\L*\n* \multiply\L*\Nhalfperiods
\divide\L*57\N*\L* \divide\N*1000\ifnum\N*=0\N*1\fi
\d**=#2\Lengthunit \d*\d** \divide\d*57\multiply\d*\n*
\r*\n*  \divide\r*\N* \ifnum\r*<2\r*2\fi
\m**\r* \divide\m**2 \l**\r* \advance\l**-\m** \N*\n* \divide\N*\r*
\dt*\d* \divide\dt*\N* \dt*.5\dt* \dt*#1\dt*
\divide\dt*1000\multiply\dt*\magnitude
\k**\r* \multiply\k**\N* \dn*\n* \advance\dn*-\k** \divide\dn*2\relax
\r*\l** \divide\r*2\advance\dn*\r* \advance\N*-1\k**#3\relax
{\rotate(#3)\mov(#2,0){\sm*}}\advance\k**\dn*
{\rotate(\k**)\mov(#2,0){\sl*}}\advance\k**-\m**\advance\l**\m**\loop\dt*-\dt*
\d*\d** \advance\d*\dt*
\advance\k**\l**{\rotate(\k**)\rmov*(\d*,0pt){\sl*}}%
\advance\N*-1\ifnum\N*>0\repeat\advance\k**\m**
{\rotate(\k**)\mov(#2,0){\sl*}}{\rotate(#4)\mov(#2,0){\sl*}}}}

\def\wavearc#1(#2)[#3,#4]{{\k**=#3\n*=#4\advance\n*-\k**
\L*=4000sp\L*#2\L* \multiply\L*\n* \multiply\L*\Nhalfperiods
\divide\L*57\N*\L* \divide\N*1000\ifnum\N*=0\N*1\fi
\d**=#2\Lengthunit \d*\d** \divide\d*57\multiply\d*\n*
\r*\n*  \divide\r*\N* \ifnum\r*=0\r*1\fi
\m**\r* \divide\m**2 \l**\r* \advance\l**-\m** \N*\n* \divide\N*\r*
\dt*\d* \divide\dt*\N* \dt*.7\dt* \dt*#1\dt*
\divide\dt*1000\multiply\dt*\magnitude
\k**\r* \multiply\k**\N* \dn*\n* \advance\dn*-\k** \divide\dn*2\relax
\divide\N*4\advance\N*-1\k**#3\relax
{\rotate(#3)\mov(#2,0){\sm*}}\advance\k**\dn*
{\rotate(\k**)\mov(#2,0){\sl*}}\advance\k**-\m**\advance\l**\m**\loop\dt*-\dt*
\d*\d** \advance\d*\dt* \dd*\d** \advance\dd*1.3\dt*
\advance\k**\r*{\rotate(\k**)\rmov*(\d*,0pt){\sl*}}\relax
\advance\k**\r*{\rotate(\k**)\rmov*(\dd*,0pt){\sl*}}\relax
\advance\k**\r*{\rotate(\k**)\rmov*(\d*,0pt){\sl*}}\relax
\advance\k**\r*
\advance\N*-1\ifnum\N*>0\repeat\advance\k**\m**
{\rotate(\k**)\mov(#2,0){\sl*}}{\rotate(#4)\mov(#2,0){\sl*}}}}

\def\gmov*#1(#2,#3)#4{\rlap{\L*=#1\Lengthunit
\xL*=#2\L* \yL*=#3\L*
\rx* \gcos*\xL* \tmp* \gsin*\yL* \advance\rx*-\tmp*
\ry* \gcos*\yL* \tmp* \gsin*\xL* \advance\ry*\tmp*
\rx*=\xscale\rx* \ry*=\yscale\ry*
\xL* \the\cos*\rx* \tmp* \the\sin*\ry* \advance\xL*-\tmp*
\yL* \the\cos*\ry* \tmp* \the\sin*\rx* \advance\yL*\tmp*
\kern\xL*\raise\yL*\hbox{#4}}}

\def\rgmov*(#1,#2)#3{\rlap{\xL*#1\yL*#2\relax
\rx* \gcos*\xL* \tmp* \gsin*\yL* \advance\rx*-\tmp*
\ry* \gcos*\yL* \tmp* \gsin*\xL* \advance\ry*\tmp*
\rx*=\xscale\rx* \ry*=\yscale\ry*
\xL* \the\cos*\rx* \tmp* \the\sin*\ry* \advance\xL*-\tmp*
\yL* \the\cos*\ry* \tmp* \the\sin*\rx* \advance\yL*\tmp*
\kern\xL*\raise\yL*\hbox{#3}}}

\def\Earc(#1)[#2,#3][#4,#5]{{\k*=#2\l*=#3\m*=\l*
\advance\m*-6\ifnum\k*>\l*\relax\else\def\xscale{#4}\def\yscale{#5}\relax
{\angle**0\rotate(#2)}\gmov*(#1,0){\sm*}\loop
\ifnum\k*<\m*\advance\k*5\relax
{\angle**0\rotate(\k*)}\gmov*(#1,0){\sl*}\repeat
{\angle**0\rotate(#3)}\gmov*(#1,0){\sl*}\relax
\def\xscale{1}\def\yscale{1}\fi}}

\def\dashEarc(#1)[#2,#3][#4,#5]{{\k**=#2\n*=#3\advance\n*-1\advance\n*-\k**
\L*=1000sp\L*#1\L* \multiply\L*\n* \multiply\L*\Nhalfperiods
\divide\L*57\N*\L* \divide\N*2000\ifnum\N*=0\N*1\fi
\r*\n*  \divide\r*\N* \ifnum\r*<2\r*2\fi
\m**\r* \divide\m**2 \l**\r* \advance\l**-\m** \N*\n* \divide\N*\r*
\k**\r*\multiply\k**\N* \dn*\n* \advance\dn*-\k** \divide\dn*2\advance\dn*\*one
\r*\l** \divide\r*2\advance\dn*\r* \advance\N*-2\k**#2\relax
\ifnum\l**<6\def\xscale{#4}\def\yscale{#5}\relax
{\angle**0\rotate(#2)}\gmov*(#1,0){\sm*}\advance\k**\dn*
{\angle**0\rotate(\k**)}\gmov*(#1,0){\sl*}\advance\k**\m**
{\angle**0\rotate(\k**)}\gmov*(#1,0){\sm*}\loop
\advance\k**\l**{\angle**0\rotate(\k**)}\gmov*(#1,0){\sl*}\advance\k**\m**
{\angle**0\rotate(\k**)}\gmov*(#1,0){\sm*}\advance\N*-1\ifnum\N*>0\repeat
{\angle**0\rotate(#3)}\gmov*(#1,0){\sl*}\def\xscale{1}\def\yscale{1}\else
\advance\k**\dn* \Earc(#1)[#2,\k**][#4,#5]\loop\advance\k**\m** \r*\k**
\advance\k**\l** {\Earc(#1)[\r*,\k**][#4,#5]}\relax
\advance\N*-1\ifnum\N*>0\repeat
\advance\k**\m**\Earc(#1)[\k**,#3][#4,#5]\fi}}

\def\triangEarc#1(#2)[#3,#4][#5,#6]{{\k**=#3\n*=#4\advance\n*-\k**
\L*=1000sp\L*#2\L* \multiply\L*\n* \multiply\L*\Nhalfperiods
\divide\L*57\N*\L* \divide\N*1000\ifnum\N*=0\N*1\fi
\d**=#2\Lengthunit \d*\d** \divide\d*57\multiply\d*\n*
\r*\n*  \divide\r*\N* \ifnum\r*<2\r*2\fi
\m**\r* \divide\m**2 \l**\r* \advance\l**-\m** \N*\n* \divide\N*\r*
\dt*\d* \divide\dt*\N* \dt*.5\dt* \dt*#1\dt*
\divide\dt*1000\multiply\dt*\magnitude
\k**\r* \multiply\k**\N* \dn*\n* \advance\dn*-\k** \divide\dn*2\relax
\r*\l** \divide\r*2\advance\dn*\r* \advance\N*-1\k**#3\relax
\def\xscale{#5}\def\yscale{#6}\relax
{\angle**0\rotate(#3)}\gmov*(#2,0){\sm*}\advance\k**\dn*
{\angle**0\rotate(\k**)}\gmov*(#2,0){\sl*}\advance\k**-\m**
\advance\l**\m**\loop\dt*-\dt* \d*\d** \advance\d*\dt*
\advance\k**\l**{\angle**0\rotate(\k**)}\rgmov*(\d*,0pt){\sl*}\relax
\advance\N*-1\ifnum\N*>0\repeat\advance\k**\m**
{\angle**0\rotate(\k**)}\gmov*(#2,0){\sl*}\relax
{\angle**0\rotate(#4)}\gmov*(#2,0){\sl*}\def\xscale{1}\def\yscale{1}}}

\def\waveEarc#1(#2)[#3,#4][#5,#6]{{\k**=#3\n*=#4\advance\n*-\k**
\L*=4000sp\L*#2\L* \multiply\L*\n* \multiply\L*\Nhalfperiods
\divide\L*57\N*\L* \divide\N*1000\ifnum\N*=0\N*1\fi
\d**=#2\Lengthunit \d*\d** \divide\d*57\multiply\d*\n*
\r*\n*  \divide\r*\N* \ifnum\r*=0\r*1\fi
\m**\r* \divide\m**2 \l**\r* \advance\l**-\m** \N*\n* \divide\N*\r*
\dt*\d* \divide\dt*\N* \dt*.7\dt* \dt*#1\dt*
\divide\dt*1000\multiply\dt*\magnitude
\k**\r* \multiply\k**\N* \dn*\n* \advance\dn*-\k** \divide\dn*2\relax
\divide\N*4\advance\N*-1\k**#3\def\xscale{#5}\def\yscale{#6}\relax
{\angle**0\rotate(#3)}\gmov*(#2,0){\sm*}\advance\k**\dn*
{\angle**0\rotate(\k**)}\gmov*(#2,0){\sl*}\advance\k**-\m**
\advance\l**\m**\loop\dt*-\dt*
\d*\d** \advance\d*\dt* \dd*\d** \advance\dd*1.3\dt*
\advance\k**\r*{\angle**0\rotate(\k**)}\rgmov*(\d*,0pt){\sl*}\relax
\advance\k**\r*{\angle**0\rotate(\k**)}\rgmov*(\dd*,0pt){\sl*}\relax
\advance\k**\r*{\angle**0\rotate(\k**)}\rgmov*(\d*,0pt){\sl*}\relax
\advance\k**\r*
\advance\N*-1\ifnum\N*>0\repeat\advance\k**\m**
{\angle**0\rotate(\k**)}\gmov*(#2,0){\sl*}\relax
{\angle**0\rotate(#4)}\gmov*(#2,0){\sl*}\def\xscale{1}\def\yscale{1}}}

\newcount\CatcodeOfAtSign
\CatcodeOfAtSign=\the\catcode`\@
\catcode`\@=11
\def\@arc#1[#2][#3]{\rlap{\Lengthunit=#1\Lengthunit
\sm*\l*arc(#2.1914,#3.0381)[#2][#3]\relax
\mov(#2.1914,#3.0381){\l*arc(#2.1622,#3.1084)[#2][#3]}\relax
\mov(#2.3536,#3.1465){\l*arc(#2.1084,#3.1622)[#2][#3]}\relax
\mov(#2.4619,#3.3086){\l*arc(#2.0381,#3.1914)[#2][#3]}}}

\def\dash@arc#1[#2][#3]{\rlap{\Lengthunit=#1\Lengthunit
\d*arc(#2.1914,#3.0381)[#2][#3]\relax
\mov(#2.1914,#3.0381){\d*arc(#2.1622,#3.1084)[#2][#3]}\relax
\mov(#2.3536,#3.1465){\d*arc(#2.1084,#3.1622)[#2][#3]}\relax
\mov(#2.4619,#3.3086){\d*arc(#2.0381,#3.1914)[#2][#3]}}}

\def\wave@arc#1[#2][#3]{\rlap{\Lengthunit=#1\Lengthunit
\w*lin(#2.1914,#3.0381)\relax
\mov(#2.1914,#3.0381){\w*lin(#2.1622,#3.1084)}\relax
\mov(#2.3536,#3.1465){\w*lin(#2.1084,#3.1622)}\relax
\mov(#2.4619,#3.3086){\w*lin(#2.0381,#3.1914)}}}

\def\bezier#1(#2,#3)(#4,#5)(#6,#7){\N*#1\l*\N* \advance\l*\*one
\d* #4\Lengthunit \advance\d* -#2\Lengthunit \multiply\d* \*two
\b* #6\Lengthunit \advance\b* -#2\Lengthunit
\advance\b*-\d* \divide\b*\N*
\d** #5\Lengthunit \advance\d** -#3\Lengthunit \multiply\d** \*two
\b** #7\Lengthunit \advance\b** -#3\Lengthunit
\advance\b** -\d** \divide\b**\N*
\mov(#2,#3){\sm*{\loop\ifnum\m*<\l*
\a*\m*\b* \advance\a*\d* \divide\a*\N* \multiply\a*\m*
\a**\m*\b** \advance\a**\d** \divide\a**\N* \multiply\a**\m*
\rmov*(\a*,\a**){\unhcopy\spl*}\advance\m*\*one\repeat}}}

\catcode`\*=12

\newcount\n@ast
\def\n@ast@#1{\n@ast0\relax\get@ast@#1\end}
\def\get@ast@#1{\ifx#1\end\let\next\relax\else
\ifx#1*\advance\n@ast1\fi\let\next\get@ast@\fi\next}

\newif\if@up \newif\if@dwn
\def\up@down@#1{\@upfalse\@dwnfalse
\if#1u\@uptrue\fi\if#1U\@uptrue\fi\if#1+\@uptrue\fi
\if#1d\@dwntrue\fi\if#1D\@dwntrue\fi\if#1-\@dwntrue\fi}

\def\halfcirc#1(#2)[#3]{{\Lengthunit=#2\Lengthunit\up@down@{#3}\relax
\if@up\mov(0,.5){\@arc[-][-]\@arc[+][-]}\fi
\if@dwn\mov(0,-.5){\@arc[-][+]\@arc[+][+]}\fi
\def\lft{\mov(0,.5){\@arc[-][-]}\mov(0,-.5){\@arc[-][+]}}\relax
\def\rght{\mov(0,.5){\@arc[+][-]}\mov(0,-.5){\@arc[+][+]}}\relax
\if#3l\lft\fi\if#3L\lft\fi\if#3r\rght\fi\if#3R\rght\fi
\n@ast@{#1}\relax
\ifnum\n@ast>0\if@up\shade[+]\fi\if@dwn\shade[-]\fi\fi
\ifnum\n@ast>1\if@up\dshade[+]\fi\if@dwn\dshade[-]\fi\fi}}

\def\halfdashcirc(#1)[#2]{{\Lengthunit=#1\Lengthunit\up@down@{#2}\relax
\if@up\mov(0,.5){\dash@arc[-][-]\dash@arc[+][-]}\fi
\if@dwn\mov(0,-.5){\dash@arc[-][+]\dash@arc[+][+]}\fi
\def\lft{\mov(0,.5){\dash@arc[-][-]}\mov(0,-.5){\dash@arc[-][+]}}\relax
\def\rght{\mov(0,.5){\dash@arc[+][-]}\mov(0,-.5){\dash@arc[+][+]}}\relax
\if#2l\lft\fi\if#2L\lft\fi\if#2r\rght\fi\if#2R\rght\fi}}

\def\halfwavecirc(#1)[#2]{{\Lengthunit=#1\Lengthunit\up@down@{#2}\relax
\if@up\mov(0,.5){\wave@arc[-][-]\wave@arc[+][-]}\fi
\if@dwn\mov(0,-.5){\wave@arc[-][+]\wave@arc[+][+]}\fi
\def\lft{\mov(0,.5){\wave@arc[-][-]}\mov(0,-.5){\wave@arc[-][+]}}\relax
\def\rght{\mov(0,.5){\wave@arc[+][-]}\mov(0,-.5){\wave@arc[+][+]}}\relax
\if#2l\lft\fi\if#2L\lft\fi\if#2r\rght\fi\if#2R\rght\fi}}

\catcode`\*=11

\def\Circle#1(#2){\halfcirc#1(#2)[u]\halfcirc#1(#2)[d]\n@ast@{#1}\relax
\ifnum\n@ast>0\L*=\xscale\Lengthunit
\ifnum\angle**=0\clap{\vrule width#2\L* height.1pt}\else
\L*=#2\L*\L*=.5\L*\special{em:linewidth .001pt}\relax
\rmov*(-\L*,0pt){\sm*}\rmov*(\L*,0pt){\sl*}\relax
\special{em:linewidth \the\linwid*}\fi\fi}

\catcode`\*=12

\def\wavecirc(#1){\halfwavecirc(#1)[u]\halfwavecirc(#1)[d]}

\def\dashcirc(#1){\halfdashcirc(#1)[u]\halfdashcirc(#1)[d]}

\def\xscale{1}
\def\yscale{1}

\def\Ellipse#1(#2)[#3,#4]{\def\xscale{#3}\def\yscale{#4}\relax
\Circle#1(#2)\def\xscale{1}\def\yscale{1}}

\def\dashEllipse(#1)[#2,#3]{\def\xscale{#2}\def\yscale{#3}\relax
\dashcirc(#1)\def\xscale{1}\def\yscale{1}}

\def\waveEllipse(#1)[#2,#3]{\def\xscale{#2}\def\yscale{#3}\relax
\wavecirc(#1)\def\xscale{1}\def\yscale{1}}

\def\halfEllipse#1(#2)[#3][#4,#5]{\def\xscale{#4}\def\yscale{#5}\relax
\halfcirc#1(#2)[#3]\def\xscale{1}\def\yscale{1}}

\def\halfdashEllipse(#1)[#2][#3,#4]{\def\xscale{#3}\def\yscale{#4}\relax
\halfdashcirc(#1)[#2]\def\xscale{1}\def\yscale{1}}

\def\halfwaveEllipse(#1)[#2][#3,#4]{\def\xscale{#3}\def\yscale{#4}\relax
\halfwavecirc(#1)[#2]\def\xscale{1}\def\yscale{1}}

\catcode`\@=\the\CatcodeOfAtSign

\section{Introduction}

The study of the energy levels of simple atomic systems (muonium, positronium,
hydrogen atom, muonic hydrogen and others) with high precision plays significant role
for the check of the Standard Model and the bound state theory with very high accuracy.
The two-particle bound states represent important tool for the exactitude the values of
fundamental physical constants (the fine structure constant, the electron and muon masses,
the proton charge radius etc.) \cite{MT}. The observation of thin effects in low energy
physics of simple atoms can be considered as necessary supplement to the construction
of large particle colliders for deep penetration to the structure of elementary
particles and search of new fundamental interactions. Such atomic experiments can
improve our knowledge about elementary particle interactions on small distances
what may be reached only at very high energies \cite{Sokolov}.

The effects of strong interactions play essential role in the
energy spectrum of the muonic hydrogen just as electronic
hydrogen. On one hand, they are connected with two
electromagnetic proton form factors (electric $G_E$ and magnetic
$G_M$) describing the distributions of the electric charge and
magnetic moment. In the Lamb shift case the main contribution to
the energy spectrum of order $(Z\alpha)^4$ is determined by the
proton charge radius $r_p$ which appeares as differential
parameter of these distributions. So, the comparison of the
experimental data and theoretical value for the Lamb shift
obtained with the corrections of high order over $\alpha$ gives
the effective approach to obtain more reliable value of the
$r_p$. The measurement of the $2P-2S$ Lamb shift in $\mu p$ with
the precision $30$ ppm allows to obtain the value of the proton
charge radius which is an order of the magnitude better in the
comparison with different methods \cite{FK}. The calculation of
the nuclear structure corrections in the hyperfine splitting of
the energy levels (see
\cite{Zemach,Guerin,ZSFC,MPK,SGK,Friar,Pineda,FM3}) can be done
only on the basis of whole proton electromagnetic form factors.
Last experimental measurements of the form factors $G_E$ and
$G_M$ were carried out in Mainz 20 years ago \cite{Simon}.

On the other one, important contribution of strong interactions to the hydrogen spectrum is
connected with the proton polarizability \cite{DS,BE,FM,CFM}. It appeares already
in the one-loop amplitudes of the muon (electron) proton electromagnetic interaction
when different baryonic resonances can be produced in the intermediate states as a
result of the virtual Compton scattering on the proton. Exact calculation of such effect can
be done by means of experimental data and theoretical models for the polarized nucleon
structure functions. The proton structure and polarizability effects lead to main
theoretical uncertainty in the expressions for different energy levels including
the hyperfine splitting of the hydrogen ground state:
\begin{equation}
\Delta E_{theor}^{HFS}=E^F\left(1+\delta^{QED}+\delta^{str}+\delta^{pol}+
\delta^{HVP}\right),~~E^F=\frac{8}{3}
\alpha^4\frac{\mu_Pm_1^2m_2^2}{(m_1+m_2)^3},
\end{equation}
where $\mu_p$ is the proton magnetic moment in nuclear magnetons, $m_1$ is the muon
mass, $m_2$ is the proton mass, $\delta^{QED}$ represents the QED contribution,
$\delta^{HVP}$ is the contribution of hadronic vacuum polarization (HVP), the corrections
$\delta^{str}$ and $\delta^{pol}$ are the proton structure and polarizability
contributions. The expression (1) is valid both for the muonic and electronic hydrogen
but the exact value of these corrections is essentially different for such atoms.
The ground state hydrogen hyperfine splitting measurement was made many years ago
with very high accuracy \cite{Hellwig}:
\begin{equation}
\Delta \nu_{exp}^{HFS}(e p)=1~420~405.751~766~7(9)~kHz.
\end{equation}
Existing difference between the theory and experiment without accounting the proton
polarizability contribution can be expressed as follows \cite{EGS}:
\begin{equation}
\frac{\Delta E^{HFS}_{theor}(e~p)-\Delta E^{exp}_{HFS}(e~p)}{E^F(e~p)}=
-4.5(1.1)\times 10^{-6},
\end{equation}
This quantity contains one of the main uncertainties connected with inaccuracies
of the proton form factor determination. Dominant part of the one-loop proton structure
correction is defined by the following expression (the Zemach correction) \cite{Zemach}:
\begin{equation}
\Delta E_Z=E^F\frac{2\mu\alpha}{\pi^2}\int \frac{d{\bf p}}{({\bf p}^2+W^2)^2}
\Biggl[\frac{G_E(-{\bf p}^2)G_M(-{\bf p}^2)}{\mu_P}-1\Biggr]=E^F(-2\mu\alpha)
R_p,~W=\alpha\mu,
\end{equation}
where $\mu$ is the reduced mass of two particles, $R_p$ is the Zemach radius. In the
coordinate representation the Zemach correction (4) is determined by the contraction
of the charge $\rho_E$ and magnetic moment $\rho_M$ distributions. The Zemach radius
represents the integral characteristic of the proton structure effects in the
hyperfine splitting of the energy levels. It may be considered as new fundamental
proton parameter in the hydrogen atom. Numerical value of the Zemach contribution is
equal
\begin{equation}
\Delta E_Z=-1.362\pm0.068~ meV,
\end{equation}
where the $5\%$ estimation of the uncertainty is connected with the measurement of the
proton electromagnetic form factors \cite{Simon}.
So, the measurement of the muonic hydrogen hyperfine splitting as for the electronic hydrogen
with similar accuracy $30$ ppm as in the case of the Lamb shift can give new information
about possible value of the contributions $\delta^{str}$ and $\delta^{pol}$ \cite{BMR}.

Such experiment demands corresponding theoretical study of different order corrections
with the same precision. Analytical calculation of the hydrogen hyperfine splitting was
carried out during many years \cite{EGS,BY} and reached the accuracy $10^{-8}$. But these
calculations can not be used directly for the muonic hydrogen after the replacement the
electron mass
to the muon mass. The reason consists in the proton structure effects. Indeed
in the case of the muonic
hydrogen the dominant region of intermediate loop momenta is of order the
muon mass. So, the
calculation of higher order amplitudes with good accuracy can be based only on their
direct integration with the account of experimental data on the proton electromagnetic form
factors.

The investigation of different contributions to the energy levels
of the muonic atoms was done many years ago in Ref. \cite{BR}.
So, at present there is need for new more complete analysis of all
possible corrections in the HFS of the $\mu p$  with the declared
accuracy $30$ ppm. Main corrections of order $\alpha^5$ to the
hyperfine splitting of the $2S$ state in the $\mu p$ were studied
in Ref.\cite{KP}. They are very important for the extraction of
the Lamb shift value $2P-2S$ in the experiment. In this study we
calculate different contributions of orders $\alpha^5$ and
$\alpha^6$ to the muonic hydrogen HFS which are determined by the
effects of electromagnetic and strong interactions. The aim of
the work consists in obtaining the numerical value of the ground
state HFS in the muonic hydrogen with designated accuracy which
can serve as reliable guide for corresponding experiment. Some
basic problems of the HFS measurement in the muonic hydrogen were
discussed in Ref.\cite{Bakalov}.

\section{Effects of vacuum polarization in the one-photon interaction}

Our calculation of different energy levels of the hydrogen-like atoms are carried out on the basis
of the quasipotential approach where the two-particle bound state is described by the
Schroedinger-type equation \cite{MF85}:

\begin{eqnarray}
\left[G^f\right]^{-1}\psi_M\equiv\left(\frac{b^2}{2\mu_R}-\frac{{\bf p}^2}{2\mu_R}\right)
\psi_M({\bf p})=
\int\frac{d{\bf q}}{(2\pi)^3}V({\bf p},{\bf q},M)\psi_M({\bf q}),
\end{eqnarray}
where
\begin{displaymath}
b^2=E_1^2-m_1^2=E_2^2-m_2^2,
\end{displaymath}
$\mu_R=E_1E_2/M$ is the relativistic reduced mass, $M=E_1+E_2$ is the bound state mass.
The quasipotential of the equation (6) is constructed in the quantum electrodynamics
by the perturbative
series using projected on positive states the two-particle off mass shell
scattering amplitude $T$ at zero relative energies of the particles:

\begin{equation}
V=V^{(1)}+V^{(2)}+V^{(3)}+...,~~~~~T=T^{(1)}+T^{(2)}+T^{(3)}+...,
\end{equation}

\begin{equation}
V^{(1)}=T^{(1)},~V^{(2)}=T^{(2)}-T^{(1)}\times G^f\times T^{(1)}, ...~~.
\end{equation}

\begin{figure}
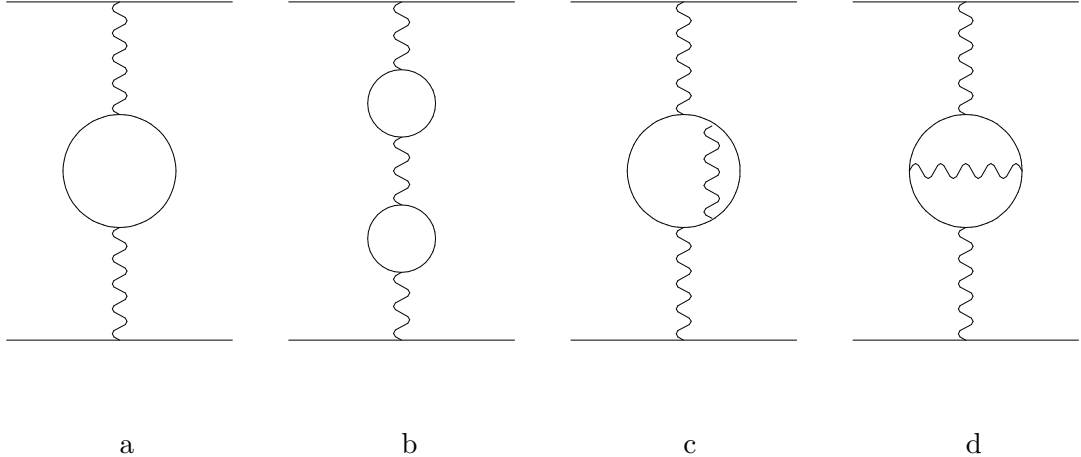

%[t!]
%\setcaptionmargin{5mm}
%\onelinecaptionsfalse
\magnitude=2000
\GRAPH(hsize=15){
\mov(0,0){\lin(2,0)}%
\mov(0,3){\lin(2,0)}%
\mov(2.5,0){\lin(2,0)}%
\mov(2.5,3){\lin(2,0)}%
\mov(5,0){\lin(2,0)}%
\mov(5,3){\lin(2,0)}%
\mov(7.5,0){\lin(2,0)}%
\mov(7.5,3){\lin(2,0)}%
\mov(1,1.5){\Circle(1.)}%
\mov(1,0){\wavelin(0,1)}%
\mov(1,2){\wavelin(0,1)}%
\mov(3.5,0){\wavelin(0,0.6)}%
\mov(3.5,1.2){\wavelin(0,0.6)}%
\mov(3.5,2.4){\wavelin(0,0.6)}%
\mov(3.5,0.9){\Circle(0.6)}%
\mov(3.5,2.1){\Circle(0.6)}%
\mov(6,0){\wavelin(0,1)}%
\mov(6,2){\wavelin(0,1)}%
\mov(6,1.5){\Circle(1.)}%
\mov(6.25,1.08){\wavelin(0,0.82)}%
\mov(8.5,0){\wavelin(0,1.)}%
\mov(8.5,2.){\wavelin(0,1)}%
\mov(8.,1.5){\wavelin(1,0)}%
\mov(8.5,1.5){\Circle(1.0)}%
\mov(1.,-1.){a}%
\mov(3.5,-1.){b}%
\mov(6.,-1.){c}%
\mov(8.5,-1.){d}%
}
\vspace{3mm}
\caption{Effects of the one- and two-loop vacuum polarization in the
one-photon interaction.}
\end{figure}

We take the ordinary Coulomb potential as initial approximation for the quasipotential
$V(\vec p,\vec q,M)$: $V(\vec p,\vec q,M)=V^C(\vec p-\vec q)+\Delta V(\vec p,\vec q,M).$

The increase of the lepton mass when we change the electronic hydrogen to the
muonic hydrogen
leads to the decrease of the Bohr radius in the $\mu p$. As a result the electron Compton wave length
and the Bohr radius are of the same order:
\begin{displaymath}
\frac{\hbar^2}{\mu e^2}:\frac{\hbar}{m_ec}=0.737384
\end{displaymath}
($m_e$ is the electron mass, $\mu$ is the reduced mass in the atom $\mu p$).
An important
consequence of last relation is the increase the role of the electron vacuum
polarization effects in the energy spectrum of the $\mu p$ \cite{t4}.
The effects of the vacuum polarization in the one-photon interaction are
shown in Fig.1.

To obtain the contribution of the diagram (a) Fig.1 (the electron vacuum polarization) to the
interaction operator there is need to make the following substitution in the photon
propagator \cite{t4}:

\begin{equation}
\frac{1}{k^2}\to \frac{\alpha}{\pi}\int_0^1 dv\frac{v^2\left(1-
\frac{v^2}{3}\right)}{k^2(1-v^2)-4m_e^2}.
\end{equation}
At $(-k^2)=\vec k^2\sim\mu_e^2(Z\alpha)^2\sim m_e^2(Z\alpha)^2$ (electronic hydrogen, $\mu_e$
is the reduced mass in hydrogen atom) we obtain $-\alpha/15\pi m_e^2$ omitting first
term in the denominator of right part of Eq.(9).
But when $\vec k^2\sim \mu^2(Z\alpha)^2\sim m_1^2(Z\alpha)^2$
(muonic hydrogen, $m_1$ is the muon mass) than $\mu \alpha$ and $m_e$ are of the same order
and it is impossible to use expansion over $\alpha$ in the denominator of Eq.(9).
To construct the hyperfine part of the quasipotential in this case
(the muonic hydrogen) in the one-photon interaction
we must use exact expression (9). We take into account that the appearance
of the electron mass $m_e$ in the denominator of the amplitude leads effectively
to the decrease the order of the correction. It is well known that the
hyperfine splitting quasipotential has the form \cite{RNF}:

\begin{equation}
V_{1\gamma}^{HFS}({\bf k})=\frac{4\pi Z\alpha}{m_1m_2}\frac{1+\kappa}{4}
\frac{1}{{\bf k}^2}[(\mathstrut\bm{\sigma}_1
\mathstrut\bm{\sigma}_2){\bf k}^2-(\mathstrut\bm{\sigma}_1
{\bf k})(\mathstrut\bm{\sigma}_2{\bf k})].
\end{equation}
For the $S$-states
\begin{equation}
V_{1\gamma}^{HFS}({\bf k})=\frac{8\pi Z\alpha}{3m_1m_2}\frac{
\mathstrut\bm{\sigma}_1
\mathstrut\bm{\sigma}_2}{4}(1+\kappa),
\end{equation}
$\kappa$=1.792847337(29) is the proton anomalous magnetic moment. Averaging
the potential (11) over the
Coulomb wave functions we obtain main contribution of order $(Z\alpha)^4$
to the HFS of the $1S$-state in the system $\mu p$ (the Fermi energy):
\begin{equation}
E^F=\frac{8}{3}(Z\alpha)^4\frac{\mu^3}{m_1m_2}(1+\kappa)=182.443~meV.
\end{equation}

The modification of the Coulomb potential due to the vacuum polarization (VP)
is determined by means of Eq.(9) in the momentum representation
as follows \cite{t4}:
\begin{equation}
V^C_{VP}({\bf k})=-4\pi Z\alpha\frac{\alpha}{\pi}\int_1^\infty\frac{\sqrt
{\xi^2-1}}{3\xi^4}\frac{(2\xi^2+1)}{{\bf k}^2+4m_e^2\xi^2}d\xi
\end{equation}
In the coordinate representation we obtain:
\begin{equation}
V^C_{VP}(r)=\frac{\alpha}{3\pi}\int_1^\infty d\xi\frac{\sqrt{\xi^2-1}(
2\xi^2+1)}{\xi^4}\left(-\frac{Z\alpha}{r}e^{-2m_e\xi r}\right).
\end{equation}
The contribution of the electron vacuum polarization to the hyperfine splitting part
of the $1\gamma$ quasipotential for the $S$-states can be derived in a similar way in
the momentum and coordinate representations:
\begin{equation}
V_{1\gamma,~VP}^{HFS}({\bf k})=\frac{4\pi Z\alpha}{m_1m_2}
\frac{(1+\kappa)}{4}\frac{2}{3}
(\mathstrut\bm{\sigma}_1
\mathstrut\bm{\sigma}_2){\bf k}^2\frac
{\alpha}{\pi}\int_1^\infty\frac{\sqrt{\xi^2-1}(2\xi^2+1)}{3\xi^4({\bf k}^2+4
m_e^2\xi^2)}d\xi,
\end{equation}
\begin{equation}
V_{1\gamma,~VP}^{HFS}(r)=\frac{8Z\alpha(1+\kappa)}{3m_1m_2}\frac
{(\mathstrut\bm{\sigma}_1
\mathstrut\bm{\sigma}_2)}{4}\frac{\alpha}{\pi}\int_1^\infty\frac{\sqrt{\xi^2-
1}(2\xi^2+1)}{3\xi^4}d\xi\left[\pi\delta({\bf r})-\frac{m_e^2\xi^2}{r}e^{-2m_e\xi r}
\right].
\end{equation}

Using Eq.(16) we can obtain the electron vacuum polarization
correction of order $\alpha^5$ to the HFS in the $\mu p$. Taking
the wave function of the $1S$-state
\begin{equation}
\psi_{100}(r)=\frac{W^{3/2}}{\sqrt{\pi}}e^{-Wr},~~~W=\mu
Z\alpha,
\end{equation}
we represent this correction in the form:
\begin{equation}
\Delta E_{1\gamma,VP}{HFS}=\frac{8\mu^3(Z\alpha)^4(1+\kappa)}{3m_1m_2}
\frac{\alpha}{\pi}\frac{m_e^3}{3W^3}\int_{m_e/W}^\infty\frac{\sqrt{\frac{W^2}
{m_e^2}\xi^2-1}}{\xi^4}\left(2\frac{W^2}{m_e^2}\xi^2+1\right)d\xi\times
\end{equation}
\begin{displaymath}
\times\left[
1-\int_0^\infty e^{-r(\xi+1)/\xi}r dr\right]=0.374~meV.
\end{displaymath}
The contribution of the muon vacuum polarization (MVP) can be found by means (16) after
the substitution $m_e\to m_1$. This correction is of order $\alpha^6$ due to the reason
mentioned above. Numerical value is equal
\begin{equation}
\Delta E_{1\gamma,~MVP}^{HFS}=E^F\frac{3}{16}\frac{\mu}{m_1}Z\alpha^2=0.002~ meV.
\end{equation}
The diagrams of the two-loop electron vacuum polarization shown in Fig.1 (b,c,d) give the contributions of
the same order $\alpha^6$. The interaction operator corresponding to the loop after
loop amplitude can be obtained using the relation (9). In the coordinate representation
\begin{equation}
V_{1\gamma,~VP-VP}^{HFS}(r)=\frac{8\pi Z\alpha(1+\kappa)}{3m_1m_2}
\frac{(\mathstrut\bm{\sigma}_1
\mathstrut\bm{\sigma}_2)}{4}\left(\frac{\alpha}{\pi}\right)^2
\int_1^\infty\frac{\sqrt{\xi^2-1}(2\xi^2+1)}{3\xi^4}d\xi\times
\end{equation}
\begin{displaymath}
\times\int_1^\infty\frac{\sqrt{\eta^2-1}(2\eta^2+1)}{3\eta^4}d\eta\left[
\delta({\bf r})-\frac{m_e^2}{\pi r(\eta^2-\xi^2)}\left(\eta^4e^{-2m_e\eta r}-
\xi^4e^{-2m_e\xi r}\right)\right],
\end{displaymath}
and the contribution to the energy spectrum
\begin{equation}
\Delta E_{1\gamma,~VP-VP}^{HFS}=0.001~ meV.
\end{equation}
To calculate the contributions of the diagrams b, c in Fig.1 which are determined by the polarization
operator of the second order it is necessary to make the substitution in the photon propagator
\cite{EGS1}:
\begin{equation}
\frac{1}{k^2}\to \left(\frac{\alpha}{\pi}\right)^2\int_0^1\frac{f(v)}
{4m_e^2+k^2(1-v^2)}dv=\left(\frac{\alpha}{\pi}\right)^2\frac{2}{3}\int_0^1 dv
\frac{v}{4m_e^2+k^2(1-v^2)}\times
\end{equation}
\begin{displaymath}
\times\Biggl\{(3-v^2)(1+v^2)\left[Li_2\left(-\frac{1-v}{1+v}\right)+2Li_2
\left(\frac{1-v}{1+v}\right)+\frac{3}{2}\ln\frac{1+v}{1-v}\ln\frac{1+v}{2}-
\ln\frac{1+v}{1-v}\ln v\right]+
\end{displaymath}
\begin{displaymath}
\left[\frac{11}{16}(3-v^2)(1+v^2)+\frac{v^4}{4}\right]\ln\frac{1+v}{1-v}+
\left[\frac{3}{2}v(3-v^2)\ln\frac{1-v^2}{4}-2v(3-v^2)\ln v\right]+
\frac{3}{8}v(5-3v^2)\Biggr\}.
\end{displaymath}

To find numerical value of this correction we write the quasipotential in the
coordinate space:
\begin{equation}
\Delta V_{1\gamma,~ 2-loop~VP}^{HFS})=\frac{8\pi
Z\alpha(1+\kappa)}
{3m_1m_2}\left(\frac{\alpha}{\pi}\right)^2\int_0^1\frac{f(v)dv}{(1-v^2)}
\left[\delta({\bf r})-\frac{m_e^2}{\pi r
(1-v^2)}e^{-\frac{2m_er}{\sqrt {1-v^2}}}\right].
\end{equation}

The potential (23) gives the contribution to the HFS in the muonic hydrogen
\begin{equation}
\Delta E_{1\gamma,~ 2-loop~VP}^{HFS}=0.002~ meV.
\end{equation}
We calculate all contribution numerically and the results are presented with the accuracy 0.001 meV.

\section{Second order of the perturbation theory}

The corrections of the second order of the perturbative series in the energy spectrum are defined
by the reduced Coulomb Green function (RCGF) \cite{VP}:

\begin{equation}
\tilde G_1({\bf r}, {\bf r'})=\sum_{l,m}\tilde g_{nl}(r,r')Y_{lm}({\bf n})
Y_{lm}^\ast({\bf n'}).
\end{equation}
The radial wave function $\tilde g_{nl}(r,r')$ was obtained in
Ref.\cite{VP} as an expansion over the Laguerre polynomials. For
the $1S$ - state
\begin{equation}
\tilde g_{10}(r,r')=-4\mu^2 Z\alpha\left(\sum_{m=2}^\infty\frac{
L_{m-1}^1(x)L_{m-1}^1(x')}{m(m-1)}+\frac{5}{2}-\frac{x}{2}-\frac{x'}{2}\right)
e^{-\frac{x+x'}{2}},
\end{equation}
where $x=2\mu Z\alpha r$, $L_n^m$ are the Laguerre polynomials:
\begin{equation}
L_n^m(x)=\frac{e^xx^{-m}}{n!}\left(\frac{d}{dx}\right)^n\left(e^{-x}
x^{n+m}\right).
\end{equation}
Some terms of the quasipotential contain the $\delta(\vec r)$ so we have to
know the quantity $\tilde G_1(\vec r,0)$. The expression for the RCGF was found in this
case in Ref.\cite{KI} on the basis of the Hoestler representation for the Coulomb Green function after
the subtraction the pole term:
\begin{equation}
\tilde G_{1S}({\bf r},0)=\frac{Z\alpha\mu^2}{4\pi}\frac{2e^{-x/2}}{x}
\left[2x(\ln x+C)+x^2-5x-2\right],
\end{equation}
where $C=0.5772...$ is the Euler constant. The main contribution of order $\alpha^5$ in the second
order of the perturbation theory can be written in general form:
\begin{equation}
\Delta E_{1~SOPT}^{HFS}=\sum_{n=2}^\infty\frac{<\psi_1^c|V_{VP}^C|\psi_n^c>
<\psi_n^c|\Delta V_{1\gamma}^{HFS}|\psi_1^c>}{E_1^c-E_n^c},
\end{equation}
where $\Delta V_{1\gamma}^{HFS}\sim\delta(\vec r)$. Using the relations (14),
(28) we can present Eq.(29) as follows:
\begin{equation}
\Delta E_{1~SOPT}^{HFS}=-E^F\frac{2\alpha}{3\pi}\int_1^\infty d\xi
\frac{\sqrt{\xi^2-1}}{\xi^2}\left(1+\frac{1}{2\xi^2}\right)\times
\end{equation}
\begin{displaymath}
\times\int_0^\infty dx e^{-x(1+\frac{m_e\xi}{W})}\left[2x(\ln x+C)+x^2-5x
-2\right]=0.734~ meV.
\end{displaymath}
The contribution of order $\alpha^6$ in the second order of the perturbative
series which is determined
by the vacuum polarization can be derived from Eq.(29) changing $\Delta V_{1\gamma}^{HFS}\to
\Delta V_{1\gamma~VP}^{HFS}$. Using exact expressions  for the wave function $\psi_1^c(\vec r)$ (17)
and the RCGF (28) we write this correction
\begin{equation}
\Delta E_{2~SOPT}^{HFS}=-E^F\alpha^2\frac{m_e^2}{W^2}\frac{8}{9\pi^2}
\int_1^\infty d\xi\left(1+\frac{1}{2\xi^2}\right)\frac{\sqrt{\xi^2-1}}{\xi^2}
\times
\end{equation}
\begin{displaymath}
\times\int_1^\infty d\eta\left(1+\frac{1}{2\eta^2}\right)\frac
{\sqrt{\eta^2-1}}{\eta^2}H(\xi,\eta,\frac{m_e}{W}),
\end{displaymath}
\begin{equation}
H(\xi,\eta,\frac{m_e}{W})=\frac{1}{\left(1+\frac{m_e\xi}{W}\right)^2}
\frac{\eta^2}{\left(1+\frac{m_e\eta}{W}\right)^2}\left[\frac{1}{\frac{W}{m_e\xi}+
\frac{W}{m_e\eta}+\frac{W^2}{m_e^2\xi\eta}}-\ln\frac{\left(\frac{W}{m_e\xi}+
\frac{W}{m_e\eta}+\frac{W^2}{m_e^2\xi\eta}\right)}{\left(1+\frac{W}{m_e\xi}\right)
\left(1+\frac{W}{m_e\eta}\right)}\right]+
\end{equation}
\begin{displaymath}
+\eta^2\left[\frac{5}{2\left(1+\frac{m_e\xi}{W}\right)^2\left(1+
\frac{m_e\eta}{W}\right)^2}-
\frac{1}{\left(1+\frac{m_e\xi}{W}\right)^2\left(1+\frac{m_e\eta}{W}\right)^3}-
\frac{1}{\left(1+\frac{m_e\xi}{W}\right)^3\left(1+\frac{m_e\eta}{W}\right)^2}
\right]+
\end{displaymath}
\begin{displaymath}
+\frac{W^2}{m_e^2}\left[\frac{1}{\left(1+\frac{m_e\xi}{W}\right)^2}
\left(1-\ln\left(1+\frac{m_e\xi}{W}\right)\right)-\frac{5}{2\left(1+\frac{m_e
\xi}{W}\right)^2}+\frac{1}{\left(1+\frac{m_e\xi}{W}\right)^3}-
\frac{1}{\left(1+\frac{m_e\xi}{W}\right)}\right].
\end{displaymath}
Numerical value of this contribution is equal
\begin{equation}
\Delta E_{2~SOPT}^{HFS}=0.002~ meV.
\end{equation}
The second order of the perturbative series gives also other relativistic corrections of order $(Z\alpha)^6$
including recoil effects which were studied in Ref.\cite{BYG,KN,FM1}. Corresponding numerical data are
in the Table 1.

\section{Proton structure and vacuum polarization effects}

The proton structure corrections in the system $\mu p$ are relatively large
in the comparison with the electronic hydrogen.
In the HFS of the muonic hydrogen these corrections are defined in the
leading order by the one-loop diagrams in Fig.2.

To construct the quasipotential corresponding to these diagrams we write
the proton tensor:

\begin{figure}
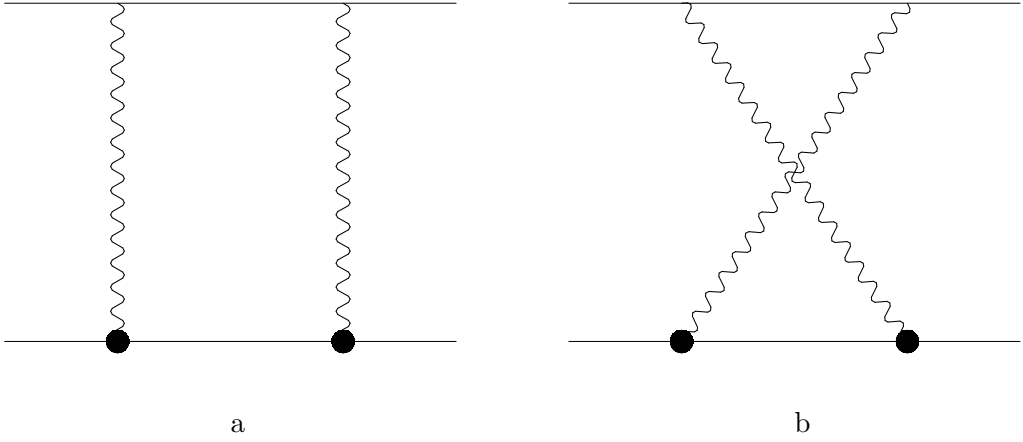

%[t!]
%\setcaptionmargin{5mm}
%\onelinecaptionsfalse
\magnitude=2000
\GRAPH(hsize=15){
\mov(0,0){\lin(1,0)}%
\mov(3,0){\lin(1,0)}%
\mov(0,3){\lin(4,0)}%
\mov(5,0){\lin(1,0)}%
\mov(8,0){\lin(1,0)}%
\mov(5,3){\lin(4,0)}%
\mov(3.,0.){\Circle**(0.2)}%
\mov(6.,0){\Circle**(0.2)}%
\mov(1.,0.){\Circle**(0.2)}%
\mov(8.,0.){\Circle**(0.2)}%
\mov(2.,-0.8){a}%
\mov(7.,-0.8){b}%
\mov(1.,0){\lin(2.,0)}%
\mov(6.,0){\lin(2.,0)}%
\mov(1,0){\wavelin(0,3)}%
\mov(3,0){\wavelin(0,3)}%
\mov(6,0){\wavelin(2,3)}%
\mov(8,0){\wavelin(-2,3)}%
}
\caption{Proton structure corrections of order $(Z\alpha)^5$. Bold circle
in the diagram represents the proton vertex operator.}
\end{figure}

\begin{equation}
M_{\mu\nu}^{(p)}=\bar u(q_2)\left[\gamma_\mu F_1+\frac{i}{2m_2}
\sigma_{\mu\omega}k^\omega F_2\right]\frac{\hat p_2-\hat k+m_2}{(p_2-k)^2-m_2^2+i0}
\left[\gamma_\nu F_1-\frac{i}{2m_2}\sigma_{\nu\lambda}k^\lambda
F_2\right]u(p_2),
\end{equation}
where $p_2, q_2$ are four momenta of the proton in initial and final states.
The construction of the potential can be essentially simplified using the projection
operators for the system muon-proton on the states with definite spin:

\begin{equation}
\hat\pi(^1S_0)=[u(p_2)\bar v(p_1)]_{S=0}=\frac{(1+\gamma^0)}{2\sqrt{2}}
\gamma_5,~~~\hat\pi(^3S_1)=[u(p_2)\bar v(p_1)]_{S=1}=\frac{(1+\gamma^0)}
{2\sqrt{2}}\hat\epsilon.
\end{equation}
where $\epsilon^\mu$ is the polarization vector of the state with the spin 1. Neglecting relative
motion momenta of the particles in the initial and final states we obtain
\begin{equation}
\Delta E_{str}^{HFS}=E^F\frac{Z\alpha m_1m_2}{8\pi n^3(1+\kappa)}
\delta_{l0}\int\frac{id^4k}{\pi^2(k^2)^2}\Biggl[\frac{16k^6k_0^2}{m_2^2}F_2^2+
\frac{32k^8}{m_2^2}F_2^2-64k^2k_0^4F_2^2+
\end{equation}
\begin{displaymath}
+16k^4k_0^2F_1^2+128k^4k_0^2F_1F_2+64k^4k_0^2F_2^2+32k^6F_1^2+64k^6F_1F_2
\Biggr]\frac{1}{(k^4-4m_1^2k_0^2)(k^4-4m_2^2k_0^2)}.
\end{displaymath}
Transforming the integration in Eq.(36) to the Euclidean space
\begin{equation}
\int d^4k=4\pi\int k^3dk\int\sin^2\phi d\phi,~~k_0=k\cos\phi,
\end{equation}
we make analytical integration over the angle $\phi$ and present the correction (36)
as one dimensional integral over the variable k:
\begin{equation}
\Delta E_{str}^{HFS}=-E^F\frac{Z\alpha}{8\pi n^3(1+\kappa)}\delta_{l0}
\int_0^\infty\frac{dk}{k}V(k),
\end{equation}
\begin{displaymath}
V(k)=\frac{2F_2^2k^2}{m_1m_2}+\frac{\mu}{(m_1-m_2)k(k+\sqrt{4m_1^2+k^2})}
\Biggl[-128F_1^2m_1^2-128F_1F_2m_1^2+16F_1^2k^2+
\end{displaymath}
\begin{displaymath}
+64F_1F_2k^2+16F_2^2k^2+\frac{32F_2^2m_1^2k^2}{m_2^2}+\frac{4F_2^2k^4}
{m_1^2}-\frac{4F_2^2k^4}{m_2^2}\Biggr]+\frac{\mu}{(m_1-m_2)k(k+\sqrt{4m_2^2+k^2})}
\times
\end{displaymath}
\begin{displaymath}
\times\left[128F_1^2m_2^2+128F_1F_2m_2^2-16F_1^2k^2-64F_1F_2k^2-48F_2^2k^2\right].
\end{displaymath}
To cancel infrared divergence in Eq.(38) it is necessary to add the contribution
of the iteration term of the quasipotential (10) in the HFS of the $\mu p$:
\begin{equation}
\Delta E_{iter,str}^{HFS}=-<V_{1\gamma}\times G^f\times V_{1\gamma}>
_{str}^{HFS}=-\frac{64}{3}\frac{\mu^4(Z\alpha)^5(1+\kappa)}{m_1m_2\pi n^3}
\int_0^\infty\frac{dk}{k^2},
\end{equation}
where the angular brackets represent averaging of the interaction operator
over the bound state
Coulomb wave functions and the index HFS shows the hyperfine part in the interaction term
of the quasipotential (10). The sum of the expressions (38) and (39) coincides with the result of
Ref.\cite{KP}. The integration in Eqs.(38) and (39) was done by means of the parameterization of the
proton electromagnetic form factors obtained from the analysis of elastic lepton-nucleon
scattering \cite{Simon}. Numerically the proton structure correction of order $(Z\alpha)^5$
is equal
\begin{equation}
\Delta E_{str}^{HFS}+\Delta E_{iter,~str}^{HFS}=-1.215~meV
\end{equation}
Moreover, the effects of the proton structure must be taken into account carefully in the amplitudes
of higher order over $\alpha$ shown in Fig.3.

\begin{figure}
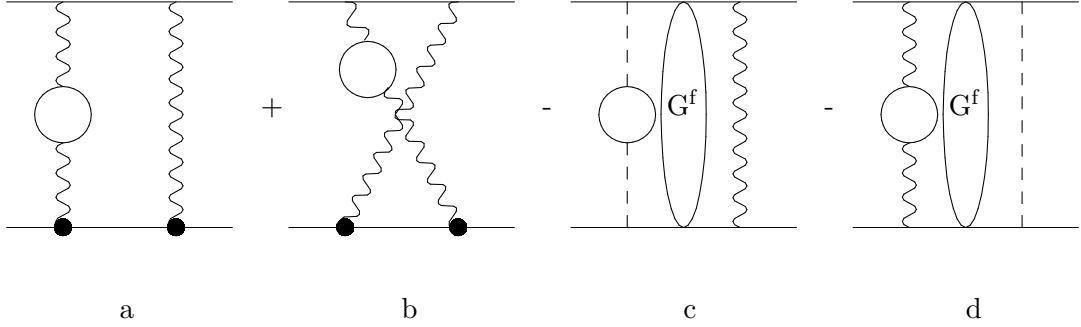

%[t!]
%\setcaptionmargin{5mm}
%\onelinecaptionsfalse
\magnitude=2000
\GRAPH(hsize=15){
\mov(0,0){\lin(2,0)}%
\mov(0,2){\lin(2,0)}%
\mov(2.5,0){\lin(2,0)}%
\mov(2.5,2){\lin(2,0)}%
\mov(5,0){\lin(2,0)}%
\mov(5,2){\lin(2,0)}%
\mov(7.5,0){\lin(2,0)}%
\mov(7.5,2){\lin(2,0)}%
\mov(0.5,0.){\Circle**(0.15)}%
\mov(1.5,0){\Circle**(0.15)}%
\mov(3.,0.){\Circle**(0.15)}%
\mov(4.,0.){\Circle**(0.15)}%
\mov(6,1){\Ellipse(0.4)[1,5]}%
\mov(5.85,1){${\rm G^f}$}%
\mov(8.5,1){\Ellipse(0.4)[1,5]}%
\mov(8.35,1){${\rm G^f}$}%
\mov(1.,-0.8){a}%
\mov(3.5,-0.8){b}%
\mov(6.,-0.8){c}%
\mov(8.5,-0.8){d}%
\mov(1.5,0){\wavelin(0,2)}%
\mov(3,0){\wavelin(1,2)}%
\mov(6.5,0){\wavelin(0,2)}%
\mov(9,0){\dashlin(0,2)}%
\mov(0.5,1.){\Circle(0.5)}%
\mov(0.5,0){\wavelin(0,0.75)}%
\mov(0.5,1.25){\wavelin(0,0.75)}%
\mov(2.25,1.){+}%
\mov(4.75,1.){-}%
\mov(7.25,1.){-}%
\mov(5.5,0){\dashlin(0,0.75)}%
\mov(5.5,1.25){\dashlin(0,0.75)}%
\mov(5.5,1.){\Circle(0.5)}%
\mov(8,0){\wavelin(0,0.75)}%
\mov(8,1.25){\wavelin(0,0.75)}%
\mov(8,1.){\Circle(0.5)}%
\mov(3,2){\wavelin(0.17,-0.34)}%
\mov(4,0){\wavelin(-0.62,1.24)}%
\mov(3.2,1.4){\Circle(0.5)}%
}
\caption{Vacuum polarization and proton structure corrections of order
$\alpha(Z\alpha)^5$. Dashed line in the diagram represents the Coulomb photon.}
\end{figure}

The contributions of the diagrams (a) and (b) in Fig.3 to the potential can be found as
for the amplitudes in Fig.2. taking into account the transformation of one exchange photon
propagator as in Eq.(9). Corresponding correction to the HFS of the energy level is equal
\begin{equation}
\Delta E_{str,VP}^{HFS}=-E^F\frac{Z\alpha}{8\pi(1+\kappa)n^3}2\frac{\alpha}
{\pi}\int_0^1\frac{v^2\left(1-\frac{v^2}{3}\right)dv}{k^2(1-v^2)+4m_e^2}\int_0
^\infty dk V_{VP}(k),
\end{equation}
where the potential $V_{VP}(k)$ differs from $V(k)$ in the relation (38) only by the factor $k^2$.
Despite of the finiteness of the integral (41) the amplitude terms of the quasipotential in Fig.3
(a), (b) must be completed by two iteration terms shown in Fig. 3 (c), (d). First addendum
$<V^c\times G^f\times\Delta V_{VP}^{HFS}>$ of order $\alpha(Z\alpha)^4$ must be subtracted
because the $2\gamma$ amplitudes (a) and (b) in Fig.3 produce lower order contribution. Second
term $<V^c_{VP}\times G^f\times V_{1\gamma}^{HFS}>$ which is also of order $\alpha(Z\alpha)^4$
has the structure similar to Eq.(29) of the second order of the perturbative
series. The contributions
of discussed iteration terms to the HFS of the $\mu p$ coincide:
\begin{equation}
\Delta E_{iter,VP+str}^{HFS}=-2<V^c\times G^f\times\Delta V_{VP}^{HFS}>^{HFS}=
-2<V_{VP}^c\times G^f\times\Delta V_{1\gamma}^{HFS}>^{HFS}=
\end{equation}
\begin{displaymath}
=-E^F\frac{4(Z\alpha)
\mu\alpha}{m_e\pi^2}\int_0^\infty dk\int_0^1\frac{v^2\left(1-\frac{v^2}{3}\right)dv}
{k^2(1-v^2)+1}.
\end{displaymath}
Numerical value of the proton structure and vacuum polarization effects
of the $2\gamma$ amplitudes
\begin{equation}
\Delta E_{VP,str}^{HFS}+2\Delta E_{iter,VP+str}^{HFS}=-0.021~meV.
\end{equation}
Hadronic vacuum polarization contribution to the HFS of the ground state
in the $\mu p$ was
studied in Ref.\cite{FM2}. Here we present it in the different form using the
expressions (38) and (41):
\begin{equation}
\Delta E_{HVP}^{HFS}=-E^F\frac{\alpha(Z\alpha)}{4\pi^2(1+\kappa)}
\int_{4m_\pi^2}^\infty\frac{\rho(s)ds}{k^2+s}\int_0^\infty dk V_{VP}(k).
\end{equation}
Dividing the integration range over s on the intervals where the cross
section of the $e^+e^-$ annihilation
into hadrons ($\rho(s)$ = $\sigma^h(e^+e^-\to hadrons)/3s\sigma_{\mu\mu}$) is known from the experiment
\cite{CMD} we can make numerical integration in Eq.(44). The result coincides with obtained in Ref.\cite{FM2}:
\begin{equation}
\Delta E_{HVP}^{HFS}=0.004~meV.
\end{equation}

\section{Proton structure effects, self energy and vertex corrections of order
$\alpha(Z\alpha)^5$}

\begin{figure}
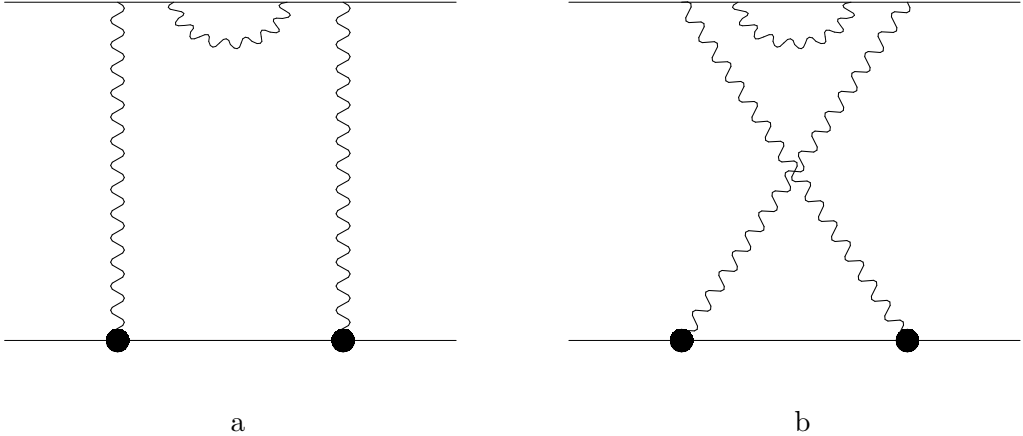

%\setcaptionmargin{5mm}
%\onelinecaptionsfalse
\magnitude=2000
\GRAPH(hsize=15){
\mov(0,0){\lin(1,0)}%
\mov(2,3){\halfwaveEllipse(1.)[D][1,0.75]}%
\mov(7,3){\halfwaveEllipse(1.)[D][1,0.75]}%
\mov(3,0){\lin(1,0)}%
\mov(0,3){\lin(4,0)}%
\mov(5,0){\lin(1,0)}%
\mov(8,0){\lin(1,0)}%
\mov(5,3){\lin(4,0)}%
\mov(3.,0.){\Circle**(0.2)}%
\mov(6.,0){\Circle**(0.2)}%
\mov(1.,0.){\Circle**(0.2)}%
\mov(8.,0.){\Circle**(0.2)}%
\mov(2.,-0.8){a}%
\mov(7.,-0.8){b}%
\mov(1.,0){\lin(2.,0)}%
\mov(6.,0){\lin(2.,0)}%
\mov(1,0){\wavelin(0,3)}%
\mov(3,0){\wavelin(0,3)}%
\mov(6,0){\wavelin(2,3)}%
\mov(8,0){\wavelin(-2,3)}%
}
\caption{Proton structure and muon self-energy effects of order
$\alpha(Z\alpha)^5$.}
\end{figure}

There exists real number of important contributions of order $\alpha^6$ which are presented
in Fig.4,5. Radiative corrections of these amplitudes including recoil effects
were studied earlier both in the Lamb shift and HFS of the hydrogen-like systems \cite{EGS,EKS,EGS2}.
Radiative photons were taken in the Fried-Yennie (FY) gauge \cite{AA,LG,FY} where the mass shell amplitudes
don't contain infrared divergences. Infrared finiteness of the Feynman diagrams in this gauge
gives the possibility to make standard subtraction on the mass shell without introducing the photon
mass. Let us consider radiative corrections which are determined by the
self-energy insertions in the muon line.
The renormalizable mass operator in the FY gauge is equal \cite{EGS}:
\begin{equation}
\Sigma^R(p)=\frac{\alpha}{\pi}(\hat p-m)^2\int_0^1 dx\frac{-3\hat px}
{m_1^2x+(m_1^2-p^2)(1-x)}.
\end{equation}

Making the insertion (46) in the lepton tensor of the two-photon exchange
diagrams and using the
projection operators (35) we can construct the hyperfine splitting part of the quasipotential
for the amplitudes in Fig. 4. In this case as before the vertex of the proton-photon interaction
is determined by electric and magnetic form factors because the typical loop momenta are of
order the muon mass. The contraction of the lepton and proton tensors over the Lorentz indices and the Dirac
$\gamma$ matrix trace calculation were made in the system Form \cite{Form}. In the Euclidean space
of the variable k we can present the correction to the HFS of the
muonic hydrogen as follows:
\begin{equation}
\Delta E_{2\gamma,SE}^{HFS}=\frac{(Z\alpha)^5\mu^3}{\pi^2 n^3}\delta_{l0}
\frac{\alpha}{\pi}\int_0^1 x dx\int_0^\infty k dk\int_0^\pi \sin^2\phi d\phi
V_{SE}(k,\phi,x),
\end{equation}
\begin{equation}
V_{SE}(k,\phi,x)=\frac{1}{(k^2+4m_2^2\cos^2\phi)[(xm_1^2+\bar xk^2)^2+
4m_1^2\bar x^2k^2\cos^2\phi]}\times
\end{equation}
\begin{displaymath}
\times\Biggl\{-\frac{4m_1^2}{m_2^2}k^2F_2^2(x+6\bar x) \cos^2\phi-\frac{8m_1^2}
{m_2^2}k^2xF_2^2+16m_1^2F_2\cos^4\phi(4F_1\bar x-F_2x-2F_2\bar x)+
\end{displaymath}
\begin{displaymath}
+16m_1^2\cos^2\phi(F_1^2x+6F_1^2\bar x+4F_1F_2x+8F_1F_2\bar x+F_2^2x+
2F_2^2\bar x)+
\end{displaymath}
\begin{displaymath}
+32m_1^2xF_1(F_1+F_2)
-\frac{4k^4}{m_2^2}F_2^2\bar x\cos^2\phi-\frac{8k^4}{m_2^2}F_2^2\bar x-
16k^2F_2^2\bar x\cos^4\phi+
\end{displaymath}
\begin{displaymath}
+16k^2\bar x\cos^2\phi(F_1^2+4F_1F_2+F_2^2)+
32k^2F_1\bar x(F_1+F_2)\Biggr\}.
\end{displaymath}
After analytical integration over the angle $\phi$ we present the contribution (47) in integral form
which was used for numerical calculation:
\begin{equation}
\Delta E_{2\gamma,SE}^{HFS}=E^F\frac{m_1m_2\alpha(Z\alpha)}{\pi^2(1\kappa)n^3}
\delta_{l0}\int_0^1 x dx\int_0^\infty dk\Biggl\{\left[-\frac{8F_2^2k^2}{m_2^2}
+32F_1(F_1+F_2)\right]\frac{1}{h_1(k,x)}+
\end{equation}
\begin{displaymath}
+\left[-\frac{k^3F_2^2}{m_2^4}-\frac{6m_1^2k^3F_2^2\bar
x}{m_2^4(xm_1^2+\bar x
k^2)}+\frac{4k}{m_2^2}(F_1^2+4F_1F_2+F_2^2)\right]
\left(\frac{1}{h_2(k,x)}-\frac{k}{h_1(k,x)}\right)+
\end{displaymath}
\begin{displaymath}
\left[\frac{2km_1^2}{m_2^2}F_2(2F_1+F_2)\bar x
-\frac{kF_2^2}{m_2^2}(xm_1^2+\bar x k^2)\right]
\left[\frac{2}{h_2^2(k,x)}-\frac{k^2}{m_2^2(xm_1^2+\bar x k^2)}\left(\frac{1}
{h_2(k,x)}-\frac{k}{h_1(k,x)}\right)\right],
\end{displaymath}
\begin{displaymath}
h_1(k,x)=k\sqrt{4m_1^2\bar x^2k^2+(xm_1^2+\bar xk^2)^2}+(xm_1^2+\bar xk^2)\sqrt
{4m_2^2+k^2},
\end{displaymath}
\begin{displaymath}
h_2(k,x)=\sqrt{4m_1^2\bar x^2k^2+(xm_1^2+\bar xk^2)^2}+(xm_1^2+\bar xk^2).
\end{displaymath}
Numerical value is equal
\begin{equation}
\Delta E_{2\gamma,SE}^{HFS}=0.008~meV.
\end{equation}

\begin{figure}
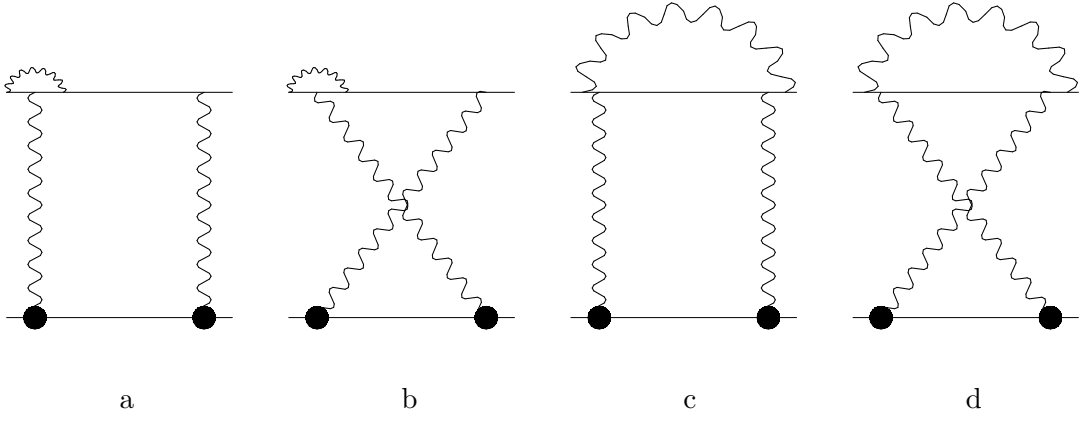

\magnitude=2000
\GRAPH(hsize=15){
\mov(0,0){\lin(2,0)}%
\mov(2.5,0){\lin(2,0)}%
\mov(2.5,2){\lin(2,0)}%
\mov(0,2){\lin(2,0)}%
\mov(5,0){\lin(2,0)}%
\mov(7.5,0){\lin(2,0)}%
\mov(7.5,2){\lin(2,0)}%
\mov(5,2){\lin(2,0)}%
\mov(0.25,0.){\Circle**(0.2)}%
\mov(0.25,0.){\wavelin(0,2)}%
\mov(1.75,0){\wavelin(0,2)}%
\mov(2.75,0){\wavelin(1.5,2)}%
\mov(4.25,0){\wavelin(-1.5,2)}%
\mov(5.25,0){\wavelin(0,2)}%
\mov(7.75,0){\wavelin(1.5,2)}%
\mov(6.75,0){\wavelin(0,2)}%
\mov(9.25,0){\wavelin(-1.5,2)}%
\mov(0.25,2){\halfwaveEllipse(0.5)[U][1.,0.8]}%
\mov(2.75,2){\halfwaveEllipse(0.5)[U][1.,0.8]}%
\mov(6.,2){\halfwaveEllipse(1.8)[U][1.,0.8]}%
\mov(8.5,2){\halfwaveEllipse(1.8)[U][1.,0.8]}%
\mov(1.75,0){\Circle**(0.2)}%
\mov(2.75,0.){\Circle**(0.2)}%
\mov(4.25,0.){\Circle**(0.2)}%
\mov(5.25,0.){\Circle**(0.2)}%
\mov(6.75,0){\Circle**(0.2)}%
\mov(7.75,0.){\Circle**(0.2)}%
\mov(9.25,0.){\Circle**(0.2)}%
\mov(1.,-0.8){a}%
\mov(3.5,-0.8){b}%
\mov(6.,-0.8){c}%
\mov(8.5,-0.8){d}%
}
\caption{Proton structure and muon vertex effects of order
$\alpha(Z\alpha)^5$.}
\end{figure}

Let us consider calculation of the vertex corrections. The renormalizable
expression of the one-particle
vertex operator in the FY gauge was obtained in Ref. \cite{ES} ($p_1^2=m_1^2$):
\begin{equation}
\Lambda_\mu^R(p,p-k)=\frac{\alpha}{4\pi}\int_0^1dx\int_0^1 dz\left[
\frac{F_\mu^{(1)}}{\Delta}+\frac{F_\mu^{(2)}}{\Delta^2}\right],
\end{equation}
where $\Delta=m_1^2x+2pk(1-x)z-k^2z(1-xz)$, the functions $F_\mu^{(1)}$, $F_\mu^{(2)}$ were
determined in Ref.\cite{ES}. The lepton tensor can be divided into two parts:
\begin{equation}
M_{\mu\nu}^{(l)(1)}=\frac{\bar v(p_1)F_\nu^{(1)}(-\hat p_1-\hat k+m_1)
\gamma_\mu v(q_1)(k^2-2k^0m_1)[m_1^2x-k^2z(1-xz)+2m_1k^2\bar x^2]}{(k^4-4k_0^2
m_1^2)[\left(m_1^2x-k^2z(1-xz)\right)^2-4m_1^2k_0^2\bar x^2z^2]},
\end{equation}
\begin{equation}
M_{\mu\nu}^{(l)(2)}=\frac{\bar v(p_1)F_\nu^{(2)}(-\hat p_1-\hat k+m_1)
\gamma_\mu v(q_1)(k^2-2k^0m_1)[m_1^2x-k^2z(1-xz)+2m_1k^2\bar x^2]^2}{(k^4-4k_0^2
m_1^2)[\left(m_1^2x-k^2z(1-xz)\right)^2-4m_1^2k_0^2\bar x^2z^2]^2}.
\end{equation}
Remaining for the simplicity the main contribution over $m_1/m_2$ we write
this type vertex corrections as follows:
\begin{equation}
\Delta E_{2\gamma,vert~1}^{HFS}=-E^F\left(\frac{\alpha}{\pi}\right)^2
\frac{8m_1m_2}{(1+\kappa)\pi n^3}\int_0^1dx\int_0^1dz\int_0\pi\sin^2\phi
d\phi \int_0^\infty kdk\times
\end{equation}
\begin{displaymath}
\times\frac{V_1(x,k,\phi)[F_1(F_1+F_2)-(1+\kappa)]}{(k^2+4m_1^2\cos^2\phi)
(k^2+4m_2^2\cos^2\phi)\left[[m_1^2x+k^2z(1-zx)]^2+4m_1^2k^2\cos^2\phi\bar
x^2z^2\right]},
\end{displaymath}
\begin{equation}
V_1(x,k,\phi)=-2m_1^4x^2(1-x)+k^2m_1^2(6x^3z^2-8x^2z^2-3x^2z+8xz-3x)+
\end{equation}
\begin{displaymath}
+k^4(4x^3z^4-6x^2z^4-5x^2z^3+12xz^3-2xz^2-6z^2+3z),
\end{displaymath}

\begin{equation}
\Delta E_{2\gamma,vert~2}^{HFS}=-E^F\left(\frac{\alpha}{\pi}\right)^2
\frac{32m_1^3m_2}{(1+\kappa)\pi
n^3}\int_0^1x(1-x)dx\int_0^1dz\int_0\pi\sin^2 \phi
d\phi\int_0^\infty k^3dk\times
\end{equation}
\begin{displaymath}
\times\frac{V_2(x,k,\phi)F_1(F_1+F_2)}{(k^2+4m_1^2\cos^2\phi)
(k^2+4m_2^2\cos^2\phi)\left[[m_1^2x+k^2z(1-zx)]^2+4m_1^2k^2\cos^2\phi\bar
x^2z^2\right]^2},
\end{displaymath}
\begin{equation}
V_2(x,k,\phi)=m_1^4x^2z(2z-1)-k^2m_1^2xz^2(4xz^2-2xz-4z+2)+
\end{equation}
\begin{displaymath}
+k^4z^3(2x^2z^3-x^2z^2-4xz^2+2xz+2z-1).
\end{displaymath}
The iteration contribution is equal
\begin{equation}
\Delta E_{iter,~2\gamma~vert}^{HFS}=<V_{1\gamma}\times G^f\times
V_{1\gamma}>_{vert}^{HFS}=F^F\left(\frac{\alpha}{\pi}\right)^2\int_0^1
dz\int_0^1dx\int_0^\infty dk\frac{4\mu}{k^2},
\end{equation}
After analytical integration in Eqs.(54) and (56) over the angle $\phi$ and subtraction (58)
(one photon is the Coulomb-like and the other one contains the hyperfine part of the potential
with the value of magnetic form factor at zero point) we have the expressions of the
diagrams (a) and (b) in Fig. 5:
\begin{equation}
\Delta E_{2\gamma,~vert}^{HFS}=-E^F\left(\frac{\alpha}{\pi}\right)^2
\int_0^1dx\int_0^1 dz\int_0^\infty dk\Biggl\{\frac{F_1(F_1+F_2)}{8k(1+\kappa)m_1^3m_2
\bar x^2z^2}\Bigl[-2m_1^4x^2\bar x+k^2m_1^2x\times
\end{equation}
\begin{displaymath}
\times(6x^2z^2-8xz^2-3xz+8z-3)+
k^4z(4x^3z^3-6x^2z^3-5x^2z^2+12xz^2-2xz-6z-3)\Bigr]\times
\end{displaymath}
\begin{displaymath}
\left[-\frac{\sqrt{1+b^2}}{b(a^2-b^2)(b^2-c^2)}+
\frac{\sqrt{1+a^2}}{a(a^2-b^2)(a^2-c^2)}+\frac{\sqrt{1+c^2}}{c(b^2-c^2)(a^2-c^2)}
\right]+\frac{F_1(F_1+F_2)x}{2(1+\kappa)m_1^3m_2k\bar x^3z^4}\times
\end{displaymath}
\begin{displaymath}
\left[m_1^4x^2z
(2z-1)-2k^2m_1^2xz^2(2xz^2-xz-2z+1)+k^4z^3(2x^2z^3-x^2z^2-4xz^2+2xz+2z-1)
\right]\times
\end{displaymath}
\begin{displaymath}
\times\Biggl[-\frac{\sqrt{1+b^2}}{b(a^2-b^2)(b^2-c^2)^2}+
\frac{\sqrt{1+a^2}}{a(a^2-b^2)(a^2-c^2)}+\frac{1}{2c\sqrt{1+c^2}(b^2-c^2)
(a^2-c^2)}-
\end{displaymath}
\begin{displaymath}
-\frac{\sqrt{1+c^2}}{2c^3(b^2-c^2)(a^2-c^2)}+
\frac{\sqrt{1+c^2}}{c(b^2-c^2)(a^2-c^2)}+\frac{\sqrt{1+c^2}}{c(b^2-c^2)(a^2-c^2)^2}
\Biggr]+\frac{4\mu}{k^2}\Biggr\},
\end{displaymath}
\begin{equation}
a^2=\frac{k^2}{4m_1^2},~~ b^2=\frac{k^2}{4m_2^2},~~ c^2=\frac{[m_1^2x+
k^2z(1-xz)]^2}{4m_1^2k^2\bar x^2z^2}.
\end{equation}
Numerical value of vertex correction (59) is equal
\begin{equation}
\Delta E_{2\gamma,~vert}^{HFS}=-0.014~meV
\end{equation}
Next vertex type diagram with one rounded photon and two exchanged photons is the diagram
of the "jellyfish" type. Its contribution to the energy spectrum is of order  $\alpha(Z\alpha)^5$.
At small loop momenta this diagram gives the finite answer in the FY gauge. The lepton tensor relating
to the diagrams (c) and (d) in Fig.5 was obtained in Ref.\cite{EGS2}:
\begin{equation}
L_{\mu\nu}^{(\mu)}=\frac{\alpha}{4\pi}\int_0^1xdx\int_0^1 (1-z)dz\sum_{n=1}^
3\frac{M_{\mu\nu}^{(n)}}{\Delta^n},
\end{equation}
where $\Delta$ has the form as in Eq.(51). The tensor functions $M_{\mu\nu}^{(n)}$ are written
explicitly in Ref.\cite{EGS2}. The character of further transformations of the amplitudes (c),
(d) in Fig.5
to construct the HFS part of the potential is the same as for other amplitudes shown in Fig.4,5.
Omitting the details of such transformations which were carried out
by means of analytical system Form \cite{Form} we write here three
contributions to the HFS corresponding to the functions $M_{\mu\nu}^{(n)}$
in the leading order over $m_1/m_2$:
\begin{equation}
\Delta E_{1,~jellyfish}^{HFS}=-\frac{64\alpha(Z\alpha)^5\mu^3\delta_{l0}}
{\pi^3n^3}\int_0^1xdx\int_0^1(1-z)(1-3xz)
\int_0^\infty kdk F_1(F_1+F_2)\times
\end{equation}
\begin{displaymath}
\times\int_0^\pi\frac{\sin^2\phi d\phi}{(k^2+4m_2^2\cos^2\phi)}
\frac{[m_1^2x+k^2z(1-xz)]}{[m_1^2x+k^2z(1-xz)]^2+4m_1^2k^2\cos^2\phi
\bar x^2z^2},
\end{displaymath}
\begin{equation}
\Delta E_{2,~jellyfish}^{HFS}=-\frac{128\alpha(Z\alpha)^5\mu^3\delta_{l0}}
{3\pi^3n^3}\int_0^1xdx\int_0^1(1-z)dz
\int_0^\infty kdk F_1(F_1+F_2)\times
\end{equation}
\begin{displaymath}
\times\int_0^\pi\frac{\sin^2\phi d\phi}{(k^2+4m_2^2\cos^2\phi)}
\frac{[m_1^2x+k^2z(1-xz)]^2[k^2xz^2(1-xz)+m_1^2(x^2z+2xz-x-3z)]}
{\left\{[m_1^2x+k^2z(1-xz)]^2+4m_1^2k^2\cos^2\phi\bar x^2z^2\right\}^2},
\end{displaymath}
\begin{equation}
\Delta E_{3,~jellyfish}^{HFS}=\frac{512\alpha(Z\alpha)^5\mu^3\delta_{l0}}
{3\pi^3n^3}\int_0^1xdx\int_0^1(1-z)z^2dz
\int_0^\infty k^3dk m_1^2F_1(F_1+F_2)\times
\end{equation}
\begin{displaymath}
\times(x+xz-x^2z-1)\int_0^\pi\frac{\sin^2\phi d\phi}{(k^2+4m_2^2\cos^2\phi)}
\frac{[m_1^2x+k^2z(1-xz)]^3}
{\left\{[m_1^2x+k^2z(1-xz)]^2+4m_1^2k^2\cos^2\phi\bar x^2z^2\right\}^3}.
\end{displaymath}
The integration over the angle $\phi$ can be done in Eqs.(63)-(65) analytically. Omitting intermediate
expressions we can write final numerical result to the HFS of the $\mu p$:
\begin{equation}
\Delta E_{jellyfish}^{HFS}=\sum_{n=1}^3\Delta E_{n,~jellyfish}^{HFS}=
0.004~meV.
\end{equation}
In the point-like proton approximation when the nucleus form factors entering the Feynman amplitudes
in Fig. 4,5 are changed on their values at $k^2=0$ ($F_1(0)=1$, $F_2(0)=\kappa$) the contributions
(63)-(65) will increase twofold.

\section{Conclusion}

\begin{table}
\caption{\label{t1}Corrections of orders $\alpha^5$, $\alpha^6$ to the ground state HFS
in the muonic hydrogen.}
\bigskip
\begin{ruledtabular}
\begin{tabular}{|c|c|c|}  \hline
Contribution to HFS of $\mu p$ & Numerical value in meV & Reference \\   \hline
The Fermi energy $E^F$ & 182.443  & \cite{EGS}, (12)   \\  \hline
Muon AMM correction $a_\mu E^F$ of order $\alpha^5,\alpha^6$ & 0.213& \cite{EGS} \\ \hline
Relativistic correction $\frac{3}{2}(Z\alpha)^2 E^F$ of order $\alpha^6$ & 0.015 &\cite{Breit}  \\  \hline
Relativistic and radiative recoil corrections&     &        \\
with the account proton AMM of order $\alpha^6$ &   0.014 &\cite{BYG}  \\   \hline
One-loop electron vacuum polarization &    &     \\
contribution of $1\gamma$ interaction of order $\alpha^5$ & 0.374& (18)  \\  \hline
One-loop muon vacuum polarization &    &   \\
contribution of $1\gamma$ interaction of order $\alpha^6$& 002& (19) \\  \hline
Vacuum polarization corrections of orders $\alpha^5$, $\alpha^6$ &    &    \\
in the second order of perturbative series   & 0.736 &(30)+(33)
\\  \hline Proton structure corrections of order $\alpha^5$ &
-1.215 &\cite{KP},~(40) \\   \hline Proton structure corrections
of order $\alpha^6$ & -0.014 & \cite{SGK}  \\ \hline
Electron vacuum polarization contribution+&     &     \\
proton structure corrections of order $\alpha^6$   &   -0.021 & (43)\\  \hline
Two-loop electron vacuum polarization &     &    \\
contribution of $1\gamma$ interaction of order $\alpha^6$  & 0.003 & (21)+(24)\\  \hline
Muon self energy + proton structure&    &    \\
correction of order $\alpha^6$   &  0.008 & (50) \\   \hline
Vertex corrections + proton structure&    &   \\
corrections of order $\alpha^6$   &  -0.014  & (61)\\  \hline
"Jellyfish" diagram correction + &   &    \\
proton structure corrections of order $\alpha^6$ &  0.004  &(66) \\  \hline
HVP contribution of order $\alpha^6$ &  0.004 & (45)\\  \hline
Proton polarizability contribution of order $\alpha^5$ & 0.084 & \cite{CFM}\\   \hline
Weak interaction contribution & 0.002 &  \cite{Eides} \\ \hline
Summary contribution &  182.638 $\pm$ 0.062 & \\   \hline
\end{tabular}
\end{ruledtabular}
\end{table}

We made the calculation of different quantum electrodynamical effects, effects of the proton structure
and polarizability, the hadron vacuum polarization to HFS of muonic hydrogen. The corrections of order
$\alpha^5$ and $\alpha^6$ were considered. Working with the vacuum polarization diagrams we take into
account that the ratio $\mu\alpha/m_e$ is very close to 1 and don't increase the order of corresponding
contributions. Obtained numerical results are presented in the Table 1. We include here also QED
corrections to the Fermi energy which are determined by muon anomalous magnetic moment $a_\mu E^F$ \cite{EGS}
(experimental value of muon anomalous magnetic moment $a_\mu^{exp}=11 659 203 (8)\times 10^{-10}$
\cite{Bennett} was used), the Breit relativistic correction of order $(Z\alpha)^6$ \cite{Breit},
relativistic and radiative recoil effects of the same order $(Z\alpha)^6m_1/m_2$ with the account
of the proton anomalous magnetic moment \cite{BYG}, the proton structure corrections of order $(Z\alpha)^6\ln (Z\alpha)^2$
\cite{SGK}, the hadron vacuum polarization contribution \cite{FM2} and the proton polarizability correction
\cite{CFM}, the weak interaction contribution due to Z boson exchange \cite{Eides}.

Let us point out some peculiarities of this investigation.

1.The effects of the vacuum polarization play very important role in the
case of the muonic hydrogen.
They lead to essential modification of the spin-dependent part of
the quasipotential of the one-photon interaction.

2. We took into account consistently the proton structure in the loop amplitudes by means
of electromagnetic form factors. The point-like proton approximation gives essentially
increased results (approximately twofold).

3. The calculation of muon self-energy and vertex corrections of order $\alpha(Z\alpha)^5$
was done on the basis of the expressions for the lepton factors in the amplitude terms of the
quasipotential obtained by Eides, Grotch and Shelyuto. We supplemented these relations by
the subtraction of the iteration terms of the potential.

Total value of the ground state HFS in the muonic hydrogen shown in the Table 1 can be considered
as definite guide for the future experiment which is prepared \cite{Bakalov}. Numerical values
of the corrections were obtained with the accuracy 0.001 meV. Theoretical error connected with
the uncertainties of fundamental physical constants (fine structure constant, the proton magnetic
moment etc.) entering the Fermi energy compose the value near $10^{-5}$ meV. Other source of
theoretical uncertainty is connected with the corrections of higher order. Its estimation can be
found from the leading correction of the next order on $\alpha$ and $m_1/m_2$ in the form:
$\alpha(Z\alpha)^2\ln(Z\alpha)^2/\pi\approx 0.0005$ meV ( the value of fine structure constant is
$\alpha^{-1}=137.03599976(50)$ \cite{MT}).

It is useful to compare the summary result for the ground state HFS in the muonic hydrogen
obtained in this work (see the Table 1) with that one which can be founded in the point like
proton approximation when we take into account only the values of the proton electromagnetic
form factors at $k^2=0$: $G_E(0)=1$, $G_M(0)=\mu$ (with the exception of the Zemach correction).
In this approximation the result of the ground state HFS may be presented with the accuracy
$O((m_1/m_2)\alpha^6)$ as follows \cite{EGS}:

\begin{equation}
\Delta E^{hfs}(QED)=E^F\Biggl\{1-2\mu\alpha R_p+\frac{3}{2}(Z\alpha)^2+a_\mu+\alpha(Z\alpha)
\left(\ln 2-\frac{5}{2}\right)+
\end{equation}
\begin{displaymath}
+\frac{1}{1+\kappa}\Biggl[-\frac{3\alpha}{\pi}\frac{m_1m_2}{m_2^2-m_1^2}\ln\frac
{m_2}{m_1}+(Z\alpha)^2\frac{\mu^2}{m_1m_2}\Bigl[\left(
2(1+\kappa)+\frac{7\kappa^2}{4}\right)\ln(Z\alpha)^{-1}-
\end{displaymath}
\begin{displaymath}
-\left(8(1+\kappa)-\frac{\kappa(12-11\kappa)}{4}\right)\ln 2+3\frac{11}{18}
+\frac{\kappa(11+31\kappa)}{36}\Bigr]\Biggr]-\frac{2}{3}(Z\alpha)^2\ln(Z\alpha)^{-2}
m_1^2r_p^2\Biggr\}=
\end{displaymath}
\begin{displaymath}
=181.177~meV.
\end{displaymath}

Essential difference between this numerical value and 182.638 meV obtained in our study can be
explained by some reasons: the modification of the Breit potential due to electron vacuum polarization
for the muonic hydrogen, the effects of the proton structure in the
two-photon and three-photon
interactions, hadronic vacuum polarization and proton polarizability effects in our calculations.
Further improvement of theoretical result presented in the Table 1 is connected first of all with
the corrections on the proton structure and polarizability which give the theoretical error near 340 ppm.
The most part of this error is determined by the proton structure corrections of order $(Z\alpha)^5$
(the Zemach correction). So, the measurement of the hyperfine splitting of the levels $1S$ and $2S$ in
the muonic hydrogen with the accuracy $30$ ppm will lead to more accurate value (with relative error $10^{-3}$)
for the Zemach radius which than can be used for the improvement of theoretical result for the
ground state hydrogen hyperfine structure and more reliable estimation of the proton polarizability
effect. The increase the number of the tasks due to excited states of simple
atoms \cite{KI2002} and the inclusion new
simple atoms where the hyperfine structure of the energy spectrum is studied
will decrease the
uncertainties in the determination of physical fundamental parameters and
increase the accuracy for the check of the Standard Model in low energy physics.

\section*{Acknowledgments}

The authors are grateful to D.D. Bakalov, A.V. Borisov, V.V. Filchenkov,
A.L. Kataev, I.B. Khriplovich, A.I. Studenikin for fruitful discussions.
Final part of this work was made during the visit of (A.P.M.) to the Humboldt University in Berlin.
He is grateful to the colleagues of the particle theory and experiment groups for warm hospitality
and German Academic Exchange Service (DAAD) for financial support of the visit under stipendium.
This work was supported in part by the Russian Fond for Basic Research
(grant No. 04-02-16085).

\end{document}